\documentclass[twocolumn,superscriptaddress,showpacs,amsmath,amssymb]{revtex4-1}

\usepackage{amsmath}
\usepackage{amssymb}
\usepackage{graphicx}
\usepackage{color}
\usepackage{epsfig}
\usepackage{amsmath}
\usepackage{amssymb}
\usepackage{epstopdf}

\DeclareMathAlphabet{\bi}{OML}{cmm}{b}{it}

\begin{document}

\title{Spin- and valley-dependent  transport through arrays of ferromagnetic silicene junctions}
\author{N. Missault}
\affiliation{Departement Fysica, Universiteit Antwerpen Groenenborgerlaan 171, B-2020 Antwerpen, Belgium}
\author{P. Vasilopoulos} 
\email{p.vasilopoulos@concordia.ca}
\affiliation{Department of Physics, Concordia University, Montreal, Quebec, Canada H3G 1M8}
\author{V. Vargiamidis} 
\email{vasileios.vargiamidis@concordia.ca}
\affiliation{Department of Physics, Concordia University, Montreal, Quebec, Canada H3G 1M8}
\author{F. M. Peeters}
\email{francois.peeters@uantwerpen.be}
\affiliation{Departement Fysica, Universiteit Antwerpen Groenenborgerlaan 171, B-2020 Antwerpen, Belgium}
\author{B. Van Duppen} 
\email{ben.vanduppen@uantwerpen.be}
\affiliation{Departement Fysica, Universiteit Antwerpen Groenenborgerlaan 171, B-2020 Antwerpen, Belgium}

\begin{abstract}
We study ballistic transport of Dirac fermions in silicene through arrays of barriers,  
of width $d$,  in the presence of an exchange field $M$  and a  tunable potential 
of height $U$ or depth $-U$. The spin- and
valley-resolved conductances as functions of $U$ or $M$, exhibit resonances
away from the Dirac point (DP) and close to it  a pronounced dip  that becomes a gap when a critical    
electric field $E_z$ is applied. This gap widens by increasing the number
of barriers and can be used to realize electric field-controlled switching of the current. 
The spin $p_s$ and valley $p_v$ polarizations of the current near the DP increase with  
$E_z$ or $M$ and can reach 100\% for certain of their
values. These field ranges widen significantly by increasing  
the number of barriers. Also,  $p_s$ and $p_v$  oscillate  nearly periodically with the separation between 
barriers or wells and  can be inverted   by reversing $M$.
\end{abstract}

\maketitle

\section{Introduction}

Silicene, a monolayer of silicon atoms forming a two-dimensional
(2D) honeycomb lattice, has been predicted to be stable \cite{guz}
and several attempts have been made to synthesize it
\cite{vog,scas1,Chen2012,Meng2013}. It has attracted considerable
attention \cite{zey} because due to its honeycomb lattice, it has Dirac cones 
similar to those of graphene but with some important  differences induced by the buckled structure of its lattice.  Contrary to
graphene, silicene has a strong intrinsic spin-orbit interaction
(SOI) which leads to a gap of approximately $1.55$ meV wide \cite{feng,falko} in the low-energy band structure. The buckled structure is a remarkable
property of silicene that graphene does not possess and can facilitate
the control \cite{falko,ezawa1} of its band gap by the application
of an external perpendicular electric field $E_z$. Accordingly,
silicene could overcome difficulties associated with potential
applications of graphene in nanoelectronics (lack of a controllable
gap) due to the available spin and valley degrees of freedom. This
and its compatibility with silicon-based technology led to ample studies
of important effects such as the spin- and valley-Hall effects
\cite{feng,barnas,tahir,tabert1}, the quantum anomalous Hall
effect \cite{ezawa1,ezawa3}, spin-valley coupling \cite{tabert2}, etc..
For a review see Ref. \onlinecite{kara}.

The strong  SOI in silicene \cite{jang}  can lead not only to
spin-resolved transport, but  also  to a cross correlation
between the valley and spin degrees of freedom. Further, silicon has
a longer spin-diffusion time \cite{mosa,quh} and spin-coherence
length \cite{sanv} compared with graphene \cite{wees}, thus making
silicene appear even more suitable for spintronics applications. Notice, for instance, the very recently reported
field-effect transistors  \cite{LT}.

In earlier works several novel features have been studied such as ferromagnetic (FM)
correlations \cite{brat} and resonant transport through double
barriers \cite{vasi7} in graphene, the conductance \cite{mmm} across FM strips on the
surface of a topological insulator or on  silicene \cite{TY,sood}. 

The study of the influence  of  electric and exchange fields on ballistic transport
through single \cite{TY, VV} and double \cite{VV} FM barriers on silicene 
led to novel results such as a field-dependent transport gap and near perfect spin and valley polarizations. 
Naturally, one wonders whether a better control can be obtained
if one uses multiple barriers or wells and how the reported effects carry over to  arrays of barriers. To our knowledge this has not been done and is the subject of this study.

The main findings of this work are as follows.
A gap develops in the charge conductance $g_c$ not only when $U$  \cite{VV} is varied but also when the strength $M$ of the FM field is;  it  widens by increasing the number of barriers.
We also quantify  the spin and valley polarizations and show that the ranges of $M$ in which they are near 100\% widen significantly by increasing the number of barriers.
In addition, we show that for wells the conductance $g_c$ oscillates with $M$
but the polarizations are much smaller than those for barriers. All these quantities oscillate nearly periodically
with the separation between barriers or wells.

The paper is organized as follows. In Sec. \ref{Sec:Calculation} we pre-\\sent
results for the spin- and valley-resolved transmission through
one or several FM junctions. 
In Sec. \ref{Sec:Results} we show that the charge conductance can
be controlled by $E_z$ or $M$ and  discuss the effects of the
field $M$ on the charge, spin, and valley transport through one or several barriers or wells. We conclude with a summary in Sec. \ref{Sec:Summary}.

\section{Transport through a FM junction}\label{Sec:Calculation}

\begin{figure}[t]
\hspace*{-0.3cm}
\includegraphics[
trim={0 0 3.5cm 0},clip,width=4.3cm]{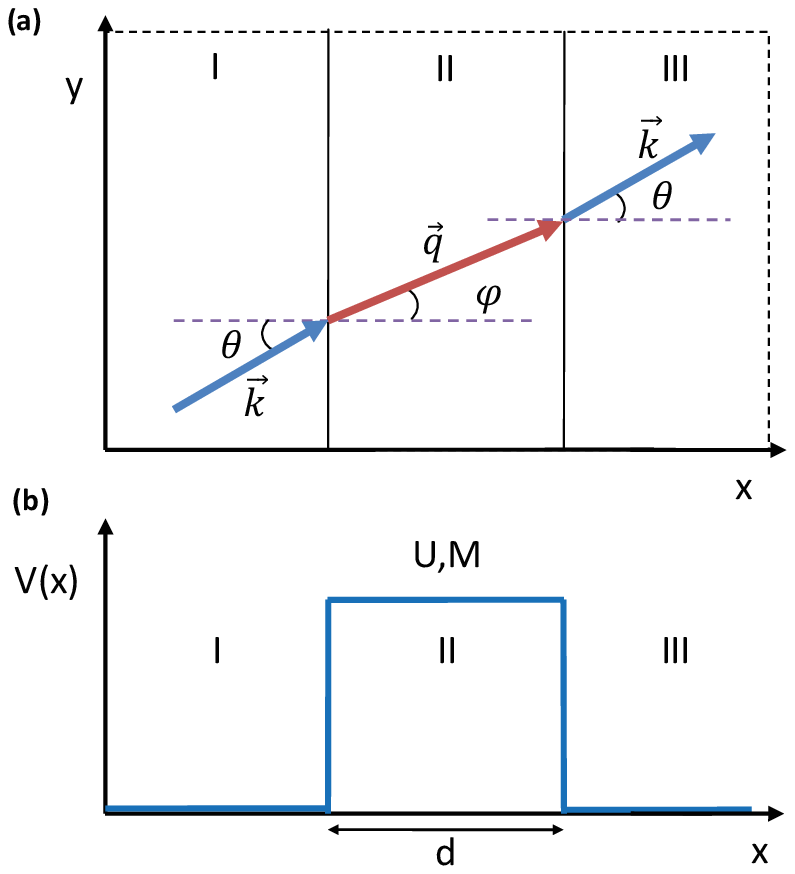}
\includegraphics[
trim={0 0 3.5cm 0},clip,width=4.3cm]{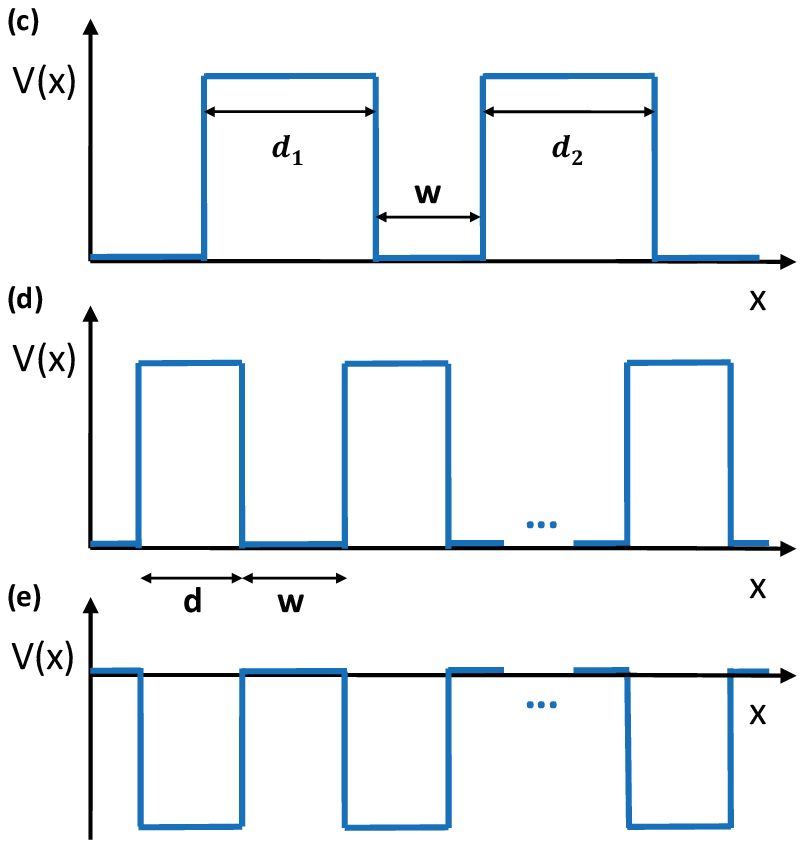}
\vspace*{-0.5cm}
\caption{(Colour online) (a) Schematics of the scattering of a  particle by
a barrier, with $\theta$ the angle of incidence and $\phi$ that of refraction. 
The vectors $\vec{k}$ and $\vec{q}$ represent the wave vectors inside (II) and outside (I, III) the barrier, respectively. (b) A single potential 
barrier:  $U$ denotes its height, $d$ its width, and $M$ the exchange field. 
(c) A double-barrier with inter-barrier separation $w$. (d), (e): Arrays of identical barriers and wells.} 
\label{Fig:fig1}
\end{figure}

We study ballistic electron transport across a FM strip in silicene with a metallic gate above it which
extends over a region of width $d$ (see Fig. \ref{Fig:fig1}(a)). The effective
Hamiltonian for low-energy fermions is given by \cite{jang}
%\color[rgb]{1,0,0}
\begin{equation}
\vspace*{-0.15cm}
\begin{aligned}
\hspace*{-0.3cm}
H_\eta &= {}  \hbar \upsilon_F \left( \tau_x k_x - \eta \tau_y k_y \right) + \Delta_{ \eta s_z} 
			 + U I - s_z M I\\ &+ \lambda_{R1}(\eta \tau_x \sigma_y-\tau_y \sigma_x)/2 + \eta a \lambda_{R2}(k_y \sigma_x - k_x \sigma_y)\tau_z.
\end{aligned}
\label{eq1}%
\end{equation}
Here $\eta = \pm 1$ distinguishes  between the  $K$ and
$K^{\prime}$  valleys, $\upsilon_F \approx 5 \times 10^5$ m/s is the Fermi velocity, and $a \approx 3.86$ {\AA} is the lattice constant.
The first term in Eq.~(\ref{eq1}) is the familiar Dirac-type Hamiltonian.  
The second term $\Delta_{ \eta s_z} =\Delta_z - \eta s_z \lambda_{so}$ describes the intrinsic SOI in
silicene through $\lambda_{so}$ and controls the SOI gap through
the perpendicular electric field term $\Delta_z = e\ell E_z$ with $2
\ell \approx 0.46$ {\AA} the vertical separation of the two
sublattices that is due to the buckled structure. The %second term, and hence the 
band gap is suppressed if the electric field is at its critical value of approximately $E_{c} \approx \lambda_{so}/e\ell \approx 17$ mV/{\AA}. The third term
represents the barrier potential due to the gate voltage, and  the
term $M I$  the exchange field due to a FM
film; $I$ is  the identity matrix. Further, $s_z = \pm 1$ represents spin-up $(\uparrow)$ and
spin-down $(\downarrow )$ states, and $\sigma_i$ and $\tau_i$ are the
Pauli matrices of respectively the spin and the sublattice pseudospin.

The first term on the second line of Eq. (1) denotes a weak {\it extrinsic} Rashba term, due to the  field $E_z$, of strength $\lambda_{R1} \propto E_z$. The buckling referred to above is also responsible for a small {\it intrinsic} Rashba effect \cite{feng} of strength $\lambda_{R2} \approx 0.7 meV$; this is the last term in Eq. (1). 
These %intrinsic and extrinsic 
Rashba terms \cite{min} result from %are consequences of 
the mixing of the $\sigma$ and $\pi$ orbitals due to atomic SOIs and, in the extrinsic case, the Stark effect. In general, they %Rashba terms 
arise when the inversion symmetry of the lattice is broken. In  silicene, however, these terms are very small and can be neglected. The extrinsic Rashba effect slightly affects the gap at the Dirac point, which closes when the  field strength is
\begin{equation}
E_c=\pm %\frac{
(\lambda_{so}/e\ell)\big(1-%\frac{
\lambda_{R1}^2/4\lambda^2_{so}\big);
\end{equation} 
the correction term $\propto \lambda_{R1}^2$ accounts  only for 0.4 {\%}. The term 
$\propto \lambda_{R2}$ % \intrinsic Rashba effect causes an
 increases the kinetic energy %given by 
by $a^2\lambda_{R2}^2k^2$ which is negligible ($10^{-9}$ {\%}) compared to the main contribution  $\hbar^2\upsilon_F^2k^2$. 
%Additionally, because the Rashba terms couple %cause a coupling between the 
%different spin-states, the electron spin is not a good quantum number \cite{ezawa3}. 
Our numerical results %in this paper 
have first been calculated using the complete four-spinor description of Eq. (1) outlined 
in the appendix.  In line though with Refs. \cite{ezawa1,ezawa3} we found no sizable differences between them and those obtained with these terms %were 
neglected. Accordingly, we  neglect them in the rest of this paper.

\begin{figure}[t]
\vspace*{-0.4cm}
\includegraphics[
width=4.2cm]{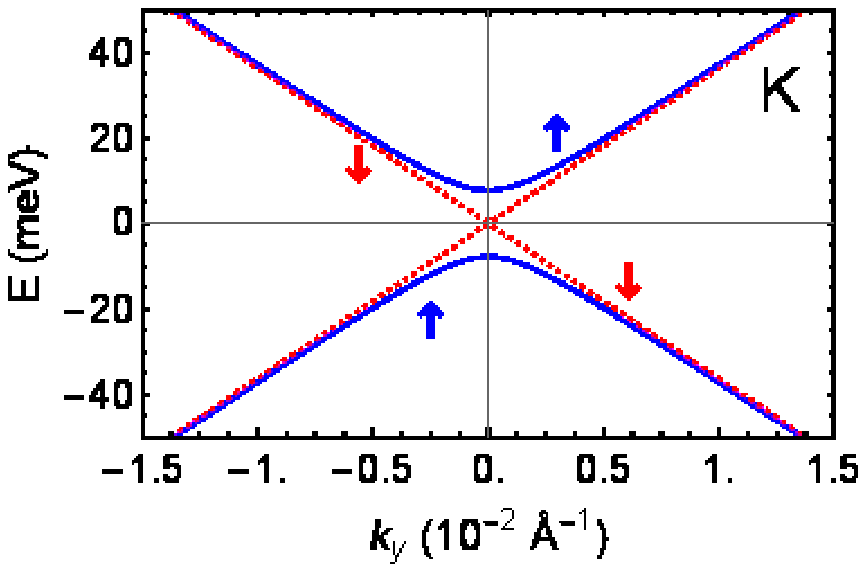}
\includegraphics[
width=4.2cm]{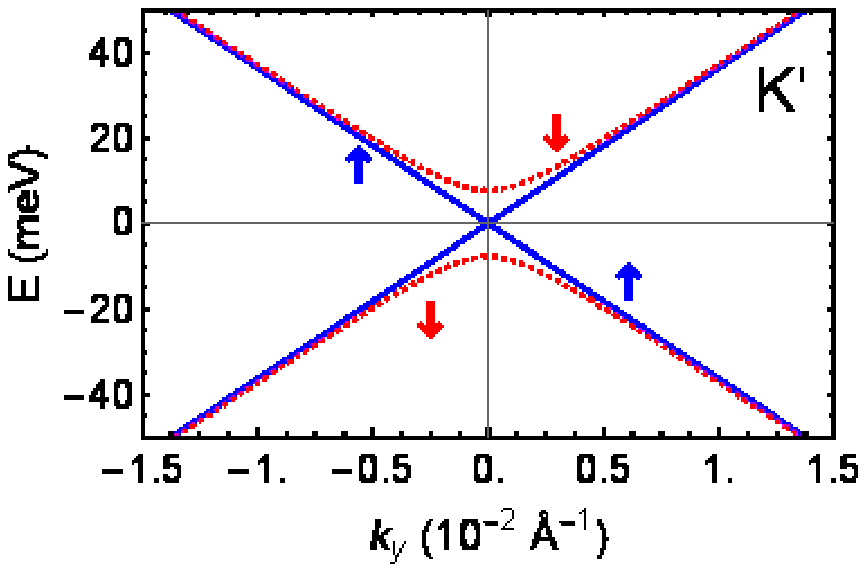}
\vspace*{-0.3cm}
\caption{(Colour online) Energy spectrum of silicene in the $K$ valley (left) and $K^{\prime}$ valley (right) of the spin up (blue curve) and spin down (red dashed curve) electrons. The applied electric field is at its critical value $E_{z} = E_{c}$. }
\label{Fig:fig2}
\end{figure}

\color[rgb]{0,0,0}
In Fig. \ref{Fig:fig2} we show the energy spectrum corresponding to
Eq. (\ref{eq1}) for $E_{z}=E_{c}$. At this value of the field  the gap of spin-down electrons closes (remains open) at the $K (K')$ valley; the inverse occurs for spin-up electrons.  
By changing $E_{z}$ and $M$ the gaps change in size and energy range respectively.
This figure suggests the polarization mechanism discussed in this work. Indeed, by tuning
$E_{z}$ and or $M$ one can make the Fermi level $E_F$ move up or down and have only one or both spin states occupied. 
This affects the propagating modes of a certain spin and valley type at 
 $E_F$ and leads to spin or valley polarization, see also Refs. \cite{TY, VV}.


\subsection{Transmission and resonances}

The eigenfunctions of Eq. (\ref{eq1}) in regions I, II, and III can be written in terms of
incident and reflected waves and are matched at the interfaces between these regions. The calculation is based on the one presented in Ref. \onlinecite{BVD} and its details are given in the appendix.  
If we neglect the very small \cite{ezawa1,ezawa3} Rashba terms %constant $\lambda_R$
, the Hamiltonian becomes block-diagonal 
and the eigenfunctions two-component spinors instead of the four-component spinors given in the appendix.  
This gives simple analytic expressions. The  transmission through a single barrier reads
\begin{equation}
T_{\eta s_z}(k_{x},k_{y})=\frac{1}{1+\sin ^{2}(dq_{x})\big[F^2(k_{x},q_{x},k_{y})-1\big]},
\label{Eq:Transmission}
\end{equation}
where
\begin{equation}
F(k_{x},q_{x},k_{y})=\frac{k_{x}^{2}\epsilon _{b}^{2}+q_{x}^{2}\epsilon
_{o}^{2}+k_{y}^{2}(\epsilon _{b}-\epsilon _{o})^{2}}{2k_{x}q_{x}\epsilon
_{b}\epsilon _{o}},
\end{equation}%
with $\epsilon _b=E-U-s_{z}M_{b}+\lambda _{so}+s_{z}\eta E_{z,b}e\ell$ and $\epsilon_{o}=E-s_{z}M_{o}+\lambda _{so}+s_{z}\eta E_{z,o}e\ell$. The subscripts $b$ and $o$ indicate  the corresponding quantities in the barrier ($b$) and outside it ($o$). Equation (\ref{Eq:Transmission}) is more general than that of
Refs. \cite{TY,VV}  because the fields   $E_z$ and $M$ are present in the entire structure whereas in Refs. \cite{TY,VV} they are present only in the barriers. It reduces to their result for $E_z=M=0$ outside the barrier. Notice  that for $q_xd=n\pi$, $n$ integer,  the transmission is perfect. These Fabry-P\'{e}rot resonances occur when half the wavelength of the wave inside the barrier fits $n$ times inside it.

A considerable simplification of Eq. (\ref{Eq:Transmission}) occurs at normal incidence ($k_y=0$):
\begin{equation}
T_{\eta s_z}(k_{x},0)=\frac{1}{1+\sin ^{2}(dq_{x})(\alpha-\alpha^{-1})^2/4},
\end{equation}
where $\alpha=k\epsilon_b/q\epsilon_o$. From this analytical result the difference with the graphene result $T=1$ is clear. Due to the SOI and the field $E_z$ the factor $\alpha$ differs from $1$. Setting $E_z=\lambda _{so}=0$, one obtains $\alpha=1$, i.e., the well-known graphene result $T=1$ no matter what the barrier width $d$ is. 
This unimpeded tunnelling at normal incidence, called Klein tunnelling, also takes place  if the electric field attains its critical value $E_{c}$. In this case, however, only the spin-up (spin-down) electrons at the K ($K^{\prime})$ point experience the Klein effect. The origin of this partial Klein tunneling can be found in Eq. (\ref{Eq:Transmission}) when 
 the condition $F^{2}(k_{x},q_{x},0)=1$ is satisfied.

The transmission probability through two barriers can also be calculated. The result is
\begin{equation}
T^{(2)}_{\eta s_z}(k_{x},k_{y})=\frac{1}{1+\sin ^{2}(dq_{x})(F^2-1)4R_{\epsilon}^{2}},
\end{equation}
with
\begin{equation}
R_{\epsilon}=\cos(w k_{x}) \cos(d q_{x})+F \sin(w k_{x})\sin(d q_{x}).
\label{Eq:AddedTwoBarriers}
\end{equation}
With $w=0$ in Eq. (6) one obtains the transmission through a single barrier, cf.  Eq. (2),  of width $2d$. 
Again the function $F$ will be responsible for Klein tunnelling at normal incidence. Because both barriers are taken 
to have the same width, single-barrier resonances are maintained. The  factor $R_{\epsilon}$, 
however, allows for an additional resonance pattern in the total double-barrier system. 

For $n$ barriers the transmission amplitude is given by
\begin{eqnarray}
t^{(n)}_{\eta s_z}& =&\big[e^{-i k_{x}l}(\cos (d q_{x})+iF\sin (w k_{x}))^{n} \nonumber \\
&-&\Theta(n-1)e^{i k_{x}l}i^{n}\sin ^{n}(d q_{x})G^{n-2}(F^{2}-1)\big]^{-1} ,
\end{eqnarray}
where $\Theta(x)$ is the Heaviside theta function; also
\begin{equation}
l =(n-1)w+(n-2)d, 
\end{equation}
\begin{equation}
G =\frac{\big[ik_{y}(\epsilon _{o}-\epsilon _{b})+k_{x}\epsilon_{b}\big]^{2}-q_{x}^{2}\epsilon _{o}^{2}}{2k_{x}q_{x}\epsilon _{b}\epsilon _{o}}.
\end{equation}
For more than two barriers the additional resonances, that occurred in the two-barrier case, fade away but the single-barrier Fabry-P\'{e}rot resonances become sharper.

\vspace*{-0.35cm}
\subsection{Conductance and  polarizations}

The spin- and valley resolved conductance is given by
\begin{eqnarray}
g_{s_z\eta } &=& g_0\int_{-\pi /2}^{\pi /2}T_{s_z\eta  }(\theta )\cos \theta, 
d\theta  
\label{Eq:ConductanceDefinition}
\end{eqnarray}
 where $g_{0}=e^{2}k_F L_y/2\pi h$, and $L_y$ is the length of the barrier along the $y$-direction. The total charge conductance $g_c$ is obtained by summing Eq. (\ref{Eq:ConductanceDefinition}) over $\eta$ and $s_z$.
 
Making use of the measurable conductance, we can define the spin polarization as
\begin{equation}
p_{s}=\frac{( g_{\uparrow K}+g_{\uparrow K^\prime }) -(g_{\downarrow K}+g_{\downarrow K^\prime }) }{g_{\uparrow K } +g_{\downarrow K}+g_{\downarrow K^\prime }+g_{\uparrow K^\prime }},
\end{equation}
with $p_{s}=1 (-1)$ if the electrons are fully polarized in the up (down) mode. Analogously, we define the valley polarization by 
\begin{equation}
p_{v}=\frac{( g_{\uparrow K }+g_{\downarrow K}) -( g_{\uparrow K^\prime }+g_{\downarrow K^\prime })}{g_{\uparrow K} +g_{\downarrow K}+g_{\downarrow
K^\prime }+g_{\uparrow K^\prime }}.
\end{equation}
Here $p_{v} = \pm 1$ corresponds to a current that is localised completely in the $K$ ($K^{\prime}$) valley. 

\vspace*{-0.4cm}

\section{Numerical results}\label{Sec:Results}

We first present results for transmission, conductance, and spin and valley polarizations through one or several  barriers and then those for one or several wells. The calculations are done using the four-component spinors given in the appendix and the fields $E_z$ and $M$ are everywhere. However, since 
the Rashba terms are %$\lambda_R$ 
indeed very  small \cite{ezawa1,ezawa3}, there is no discernible difference, in the parameter ranges used, between   the  four-component spinor results  and the analytic ones based on Eqs.(2)-(5) and (7). Accordingly, we show results  for the case these terms are neglected and  the fields $E_z$ and $M$ are inside the barriers.  

\subsection{Barriers}
\begin{figure}[t]
\hspace*{-0.4cm}
\includegraphics[
width=4.05cm,height=4.05cm%width=0.23\textwidth,clip
]{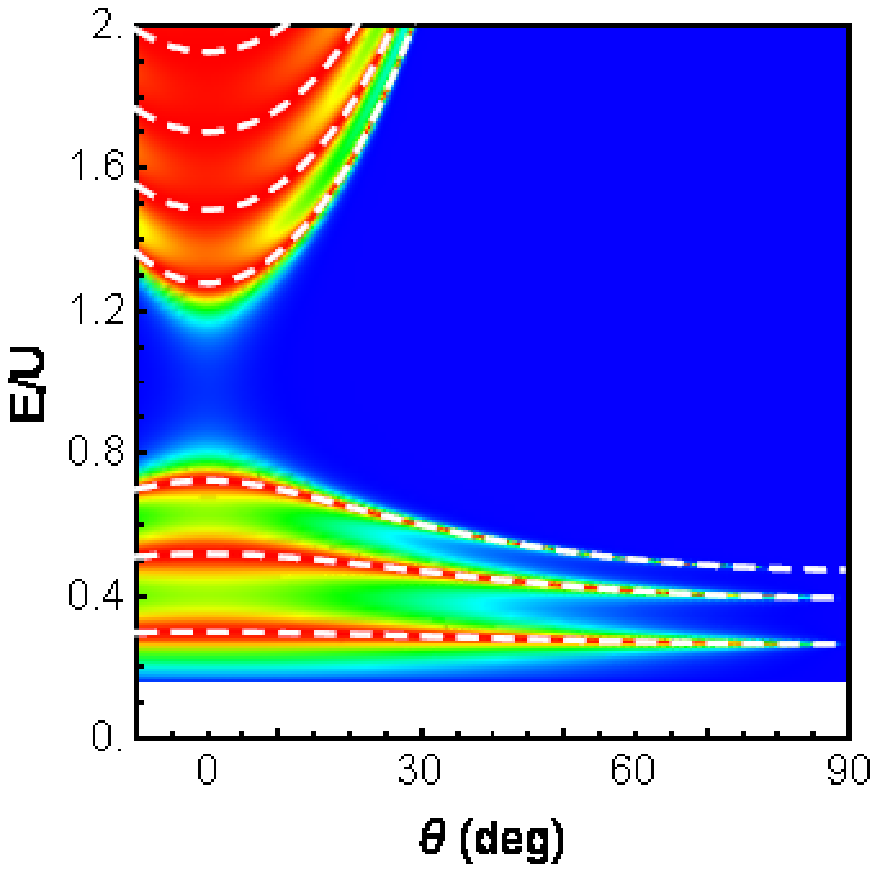}
\includegraphics[
width=3.85cm,
height=4.05cm
]{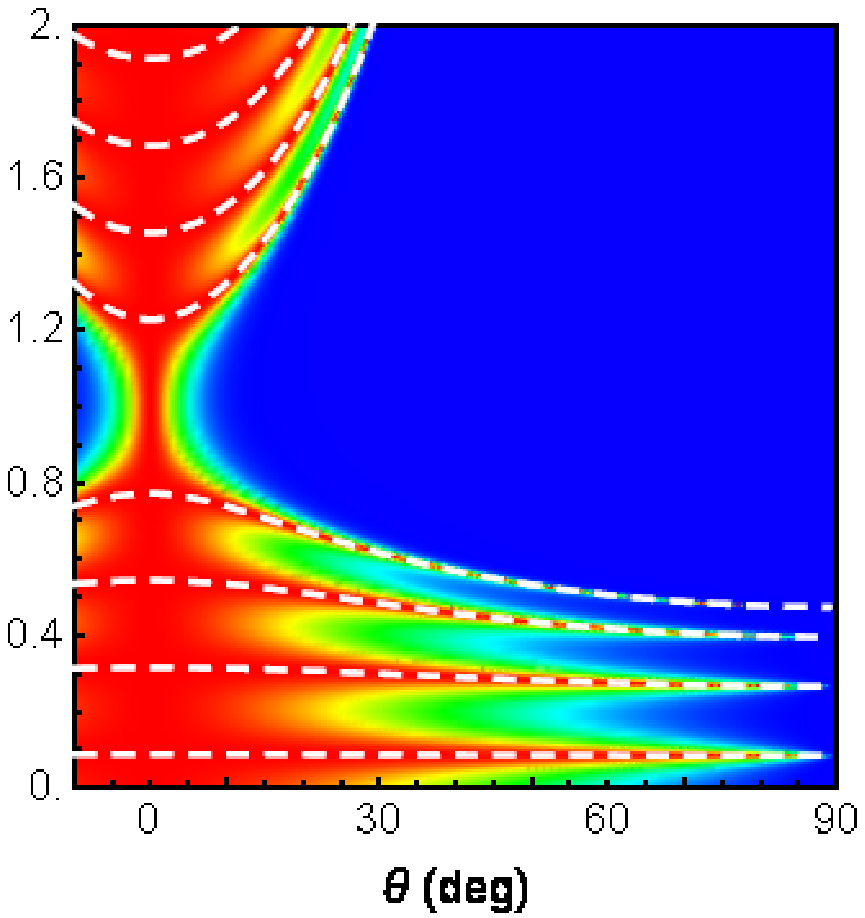}
\includegraphics[
width=0.8cm,height=4.05cm
]{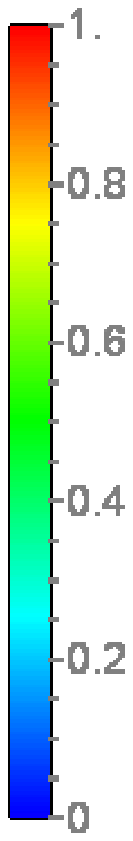} 
\hspace*{-0.4cm}
\includegraphics[
width=4.05cm,height=4.05cm
]{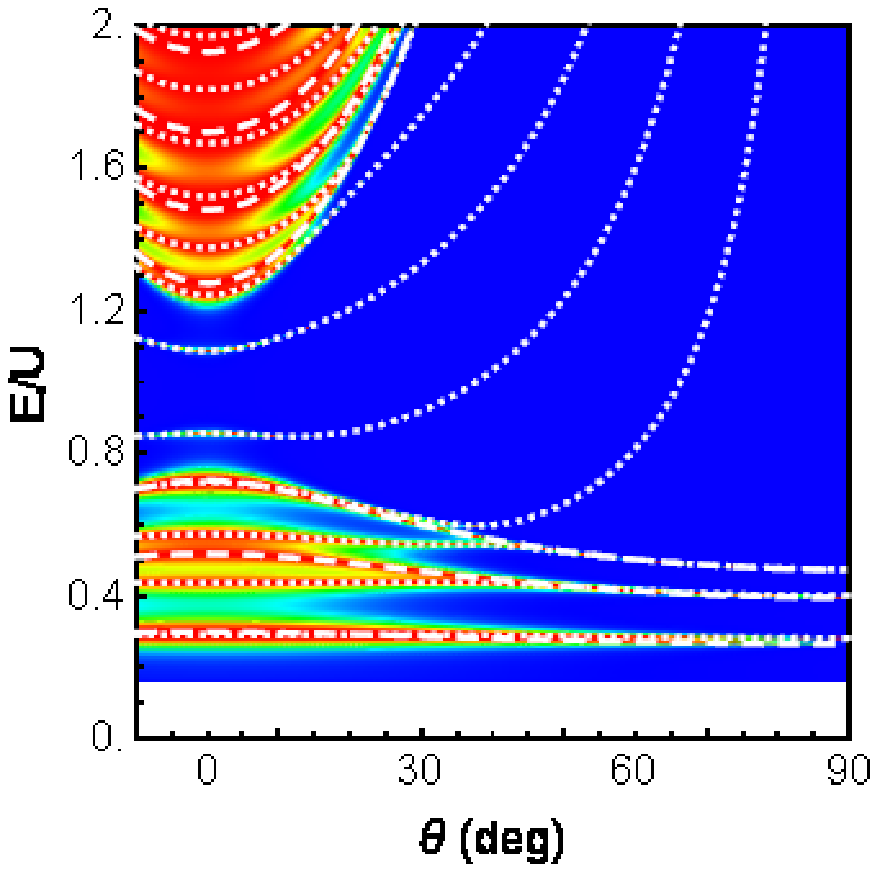}
\includegraphics[
width=3.85cm,height=4.05cm
]{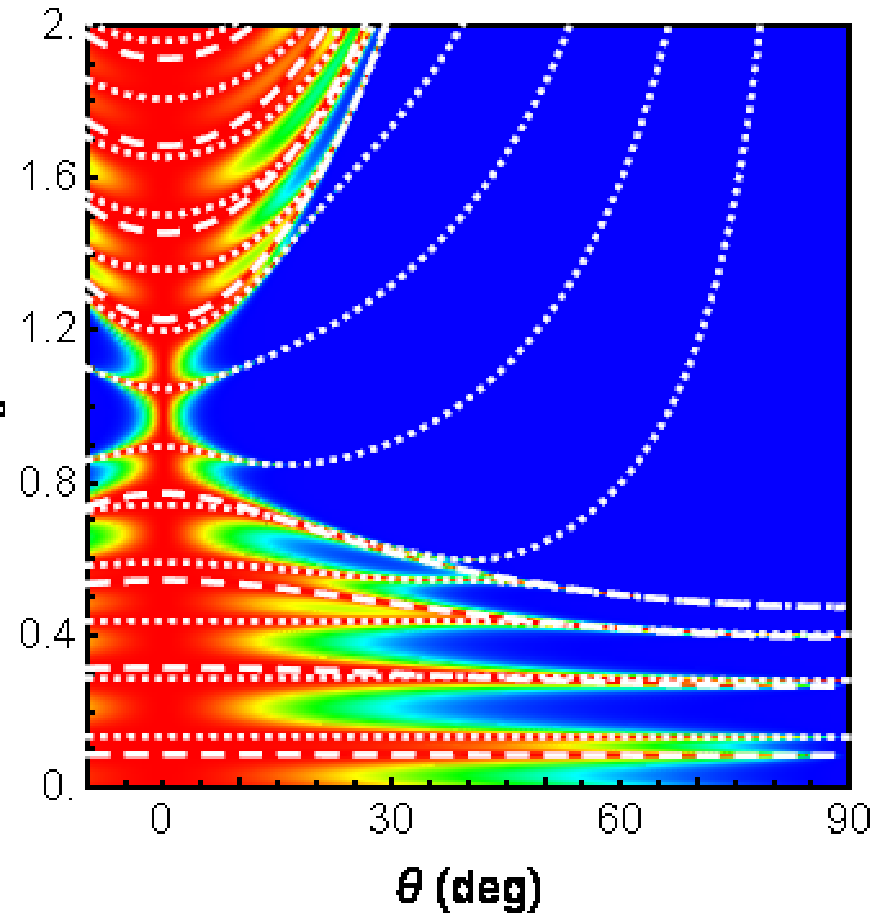}
\includegraphics[
width=0.8cm,height=4.05cm
]{Legend01.eps} 
\hspace*{-0.4cm}
\includegraphics[
width=4.05cm,height=4.05cm
]{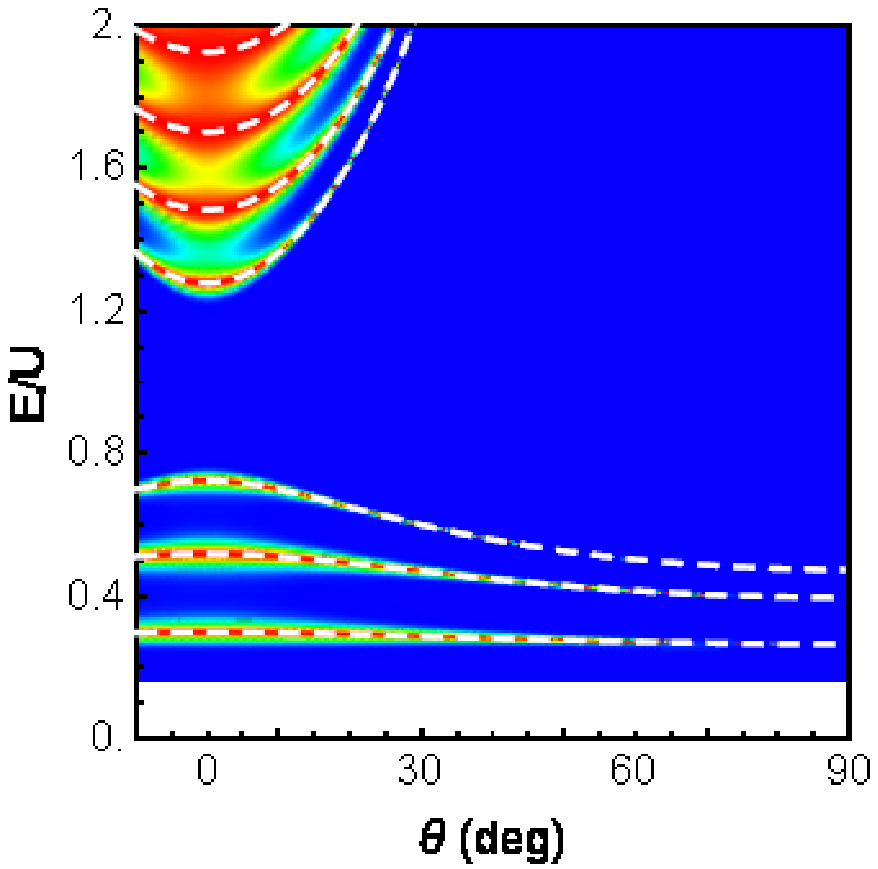}
\includegraphics[
width=3.85cm,height=4.05cm
]{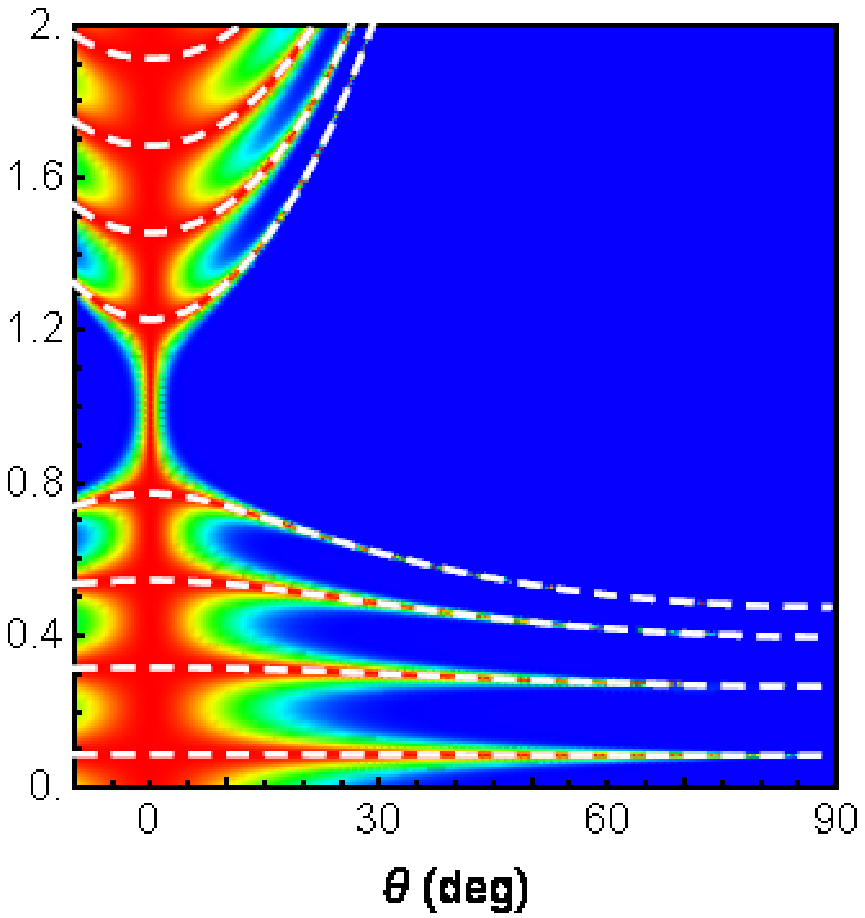}
\includegraphics[
width=0.8cm,height=4.05cm
]{Legend01.eps} 
\vspace*{-0.6cm}
\caption{(Colour online) $(E,\theta)$ contour plots of the transmission probability through $1, 2$, and $10$ barriers
in the 1$^{\text{st}}$, 2$^{\text{nd}}$, and 3$^{\text{rd}}$ row, respectively. The left (right) column is for spin-up (spin-down) electrons at the $K$ point. The dashed white curves show the resonances calculated by Eq. (\ref{Eq:Resonances}) and the dotted white curves correspond to the solutions of $R_{\epsilon} = 0$ from Eq. (\ref{Eq:AddedTwoBarriers}). In the white region in the left column the transmission is undefined due to the lack of propagating states outside the barrier.
Parameters used: $d=100$ nm, $w=50$ nm, $M=0 \lambda_{so}$ , $E_z=E_{c}$, and $U=50$ meV.}
\label{Fig:Transmission}
\end{figure}

In Fig. \ref{Fig:Transmission} we present $(E,\theta)$ contour plots of the spin resolved transmission through one, two, and ten barriers in the 1st, 2nd, and 3rd row, respectively, at the $K$ valley. The left (right) column is for spin-up (spin-down) electrons. 
The electric field is chosen to be the critical field $E_{c}$ as defined above. Accordingly, the spin-down electrons have a gapless and linear energy spectrum and therefore for normal incidence ($\theta=0$) the transmission for this state is 1 due to Klein tunneling.
Since the spin-up state's spectrum still shows a gap, there are no propagating modes in the barrier if the Fermi energy is close to $U$. Thus, %erefore 
the transmission is suppressed in an energy region of the size of the gap around the top of the barrier. 

Moving to the 2nd row of Fig. 3, for two barriers, one sees additional resonances.
Remarkably they also appear in the gap indicating that they correspond to localized states in the region between the barriers. 
Note, however, that they disappear as we move to the 3rd row (10 barriers). This is because the additional resonances appearing for $n=2$ are due to a confinement in the total barrier structure and become highly damped for $n=10$. The only resonances that are undamped in this case are the single-barrier resonances since they are shared by all barriers. 
These resonances are sharper since the near-resonance evanescent  modes that contribute to the transmission 
are suppressed because the effective length of the barrier is larger. The same mechanism also renders the gap in the transmission wider for $\theta=0$. 

The shape of the Fabry-P\'{e}rot resonances in the ($E$,$\theta$) plane can be understood by applying the condition $q_{x}d=n\pi$ for the energy. The result for the simplified case where $E_{z}=M=0$ is ($r=n\pi \upsilon_{F}/Vd$)
\begin{equation}
E(\theta )/U=
\text{sign }U\pm 
\big [1+ (r^2-1)\cos^{2}\theta)\big ]^{1/2}\big/\cos ^{2}\theta 
\label{Eq:Resonances}
\end{equation}
This relation is shown as dashed lines in Fig. \ref{Fig:Transmission}. For the values of $n$ for which $r^{2}-1$ is small, the energy solutions become approximately independent of $\theta$, which corresponds to resonances at low energies almost independent of $\theta$. For other values of $n$ the resonant energies behave proportional to $1/\cos^{2}\theta$.

\begin{figure}[t]
\includegraphics[width=4cm,height=4cm
]{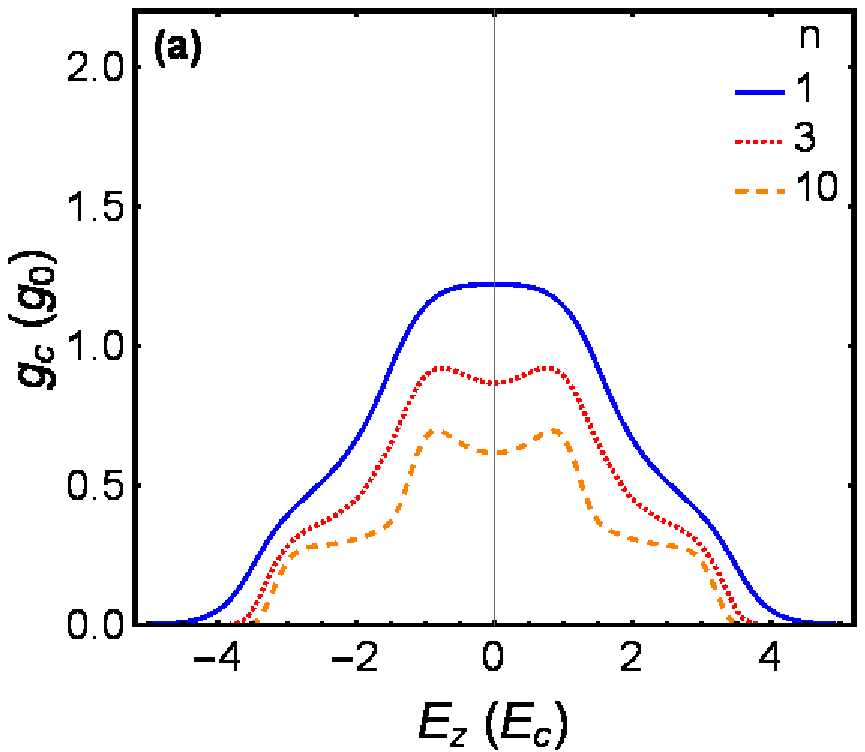}
\includegraphics[width=3.85cm,height=4cm
]{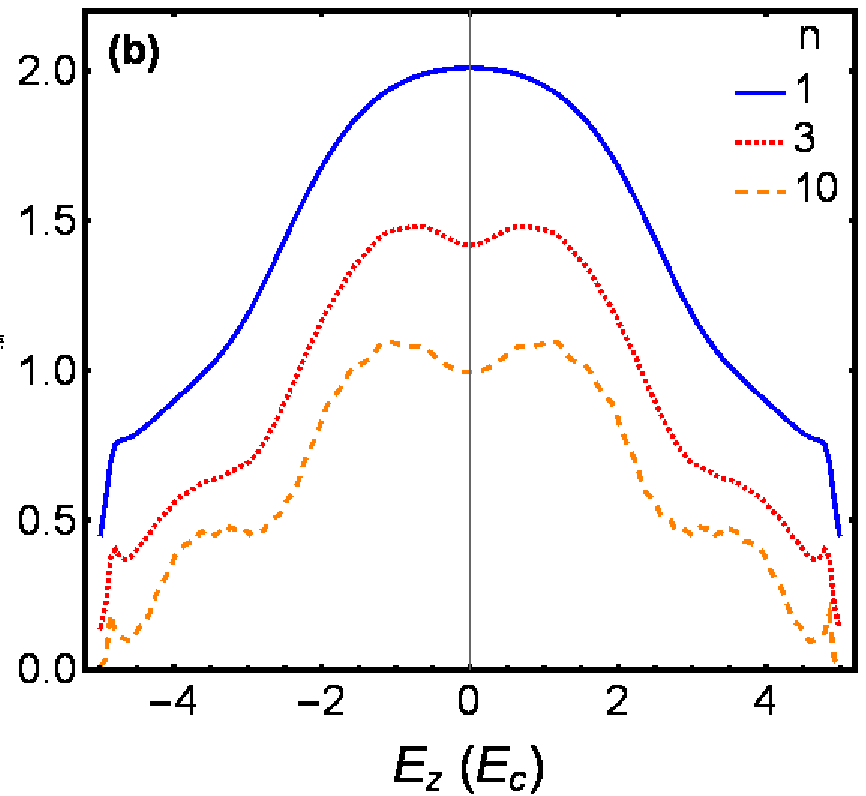}
\includegraphics[width=4.05cm,height=4cm
]{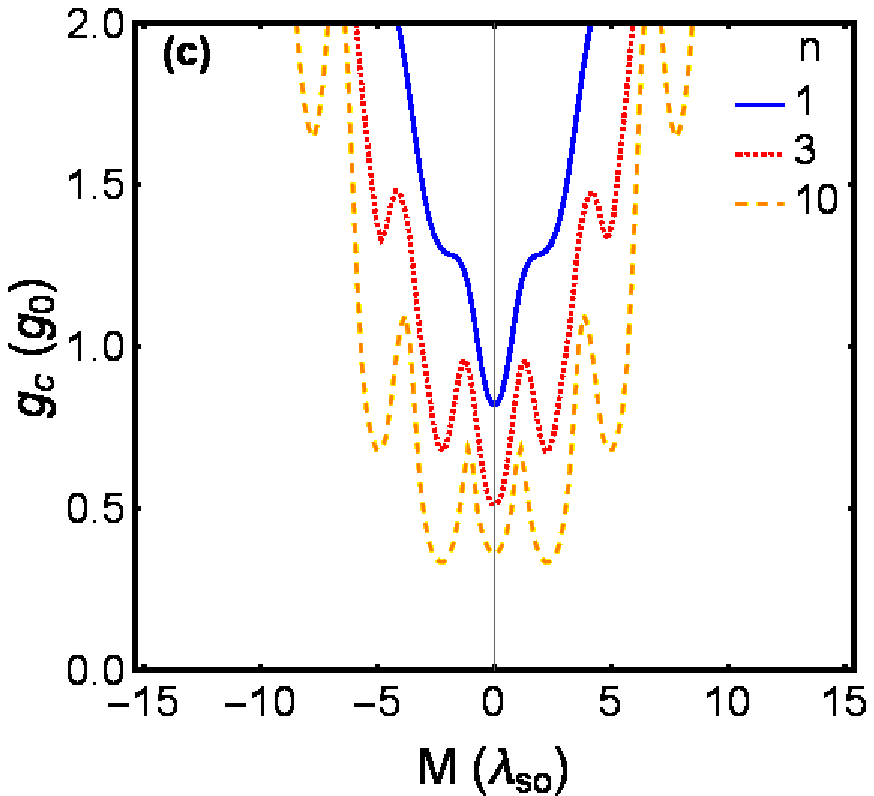}
\includegraphics[width=3.85cm,height=4cm
]{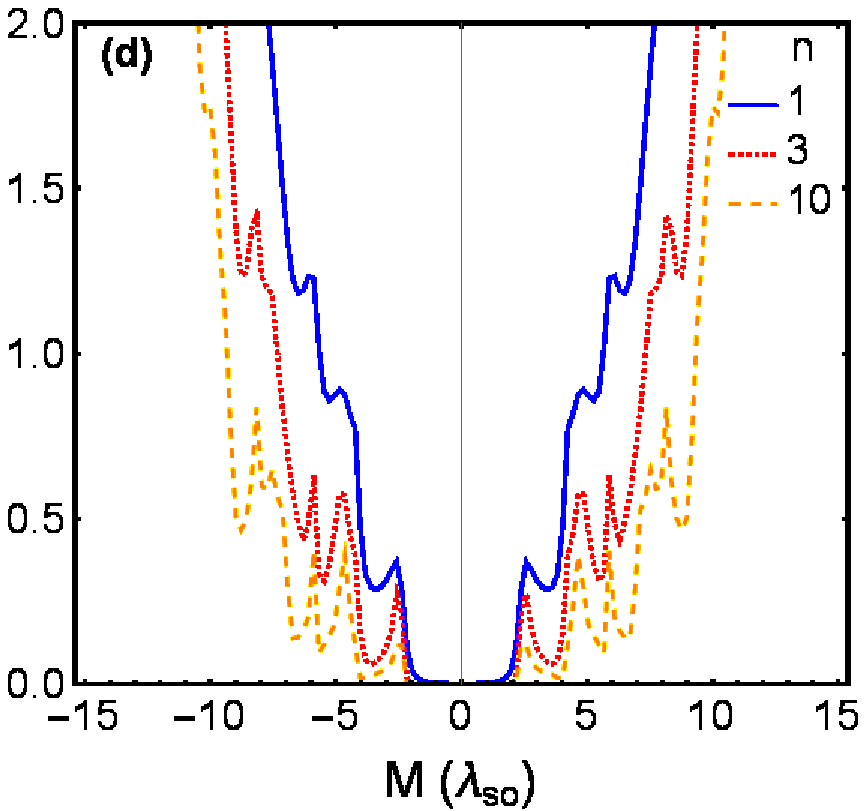}
\vspace*{-0.3cm}
\caption{(Colour online) (a) and (b): conductance through $n$ barriers versus electric field $E_z$ for $M=\lambda_{so}$ and $M=4\lambda_{so}$ respectively.
(c) and (d): Conductance versus  $M$ for $E_z=E_{c}$ and $E_z=5 E_{c}$, respectively. 
Other parameters: $d=100$ nm, $w=50$ nm, $U=50$ meV and $E=40$ meV.}
\label{Fig:ConductanceFirst}
\end{figure}

Because experimentally one measures the conductance and not the transmission, we show in Fig. \ref{Fig:ConductanceFirst} the conductance as a function of the field $E_z$ (first row) and  
of the field $M$ (second row) for $1, 3$, and $10$  barriers as indicated. 
We see that the conductance is symmetric with respect to both  $E_z$ and  $M$, it is reduced by increasing the number of barriers, and, perhaps more important, it develops a gap, i.e., it vanishes, when plotted versus $M$, for very strong fields $E_z$. The physical origin of the transport gap lies in the suppression of evanescent modes as $E_z$ increases. For one barrier this is similar to the transport gap of the conductance, plotted versus the energy at the Dirac point \cite{VV}. If $E_F$ is fixed at the correct value, there are no propagating modes inside the barrier region. The field $M$ can then be used to shift the spectra of both spins so as to coincide with $E_F$. 
This gives rise to the observed abrupt increase in the conductance as a function of $M$. With many barriers in the system, the contribution of evanescent modes near the edges of the gap is exponentially suppressed. Therefore, the gap is widened and sharpened. 
	
\begin{figure}[t]
\includegraphics[width=4.2cm,height=4cm
]{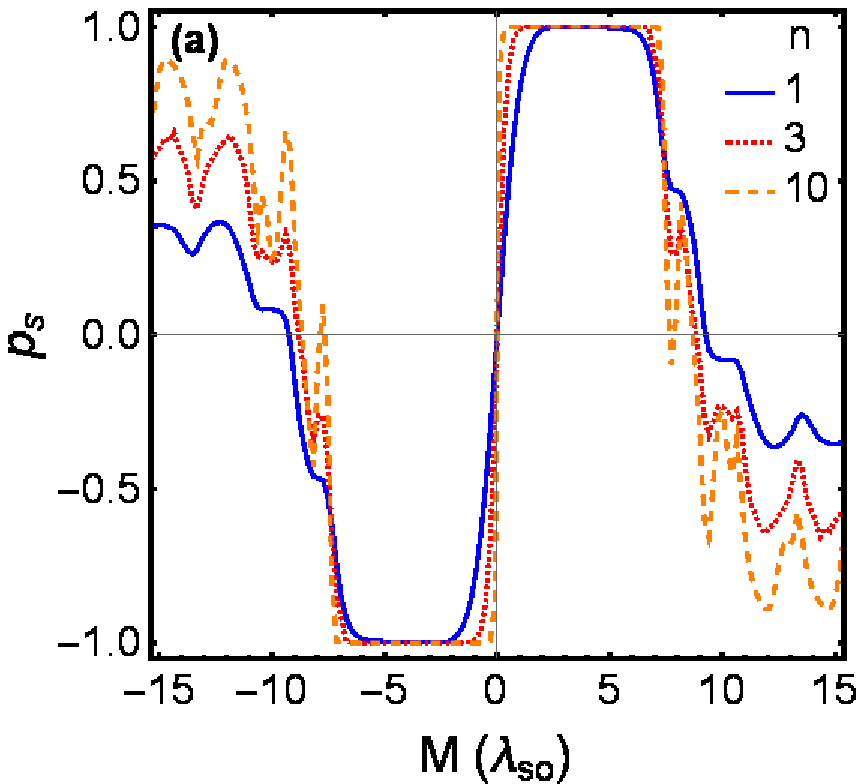}
\includegraphics[width=4.2cm,height=4cm
]{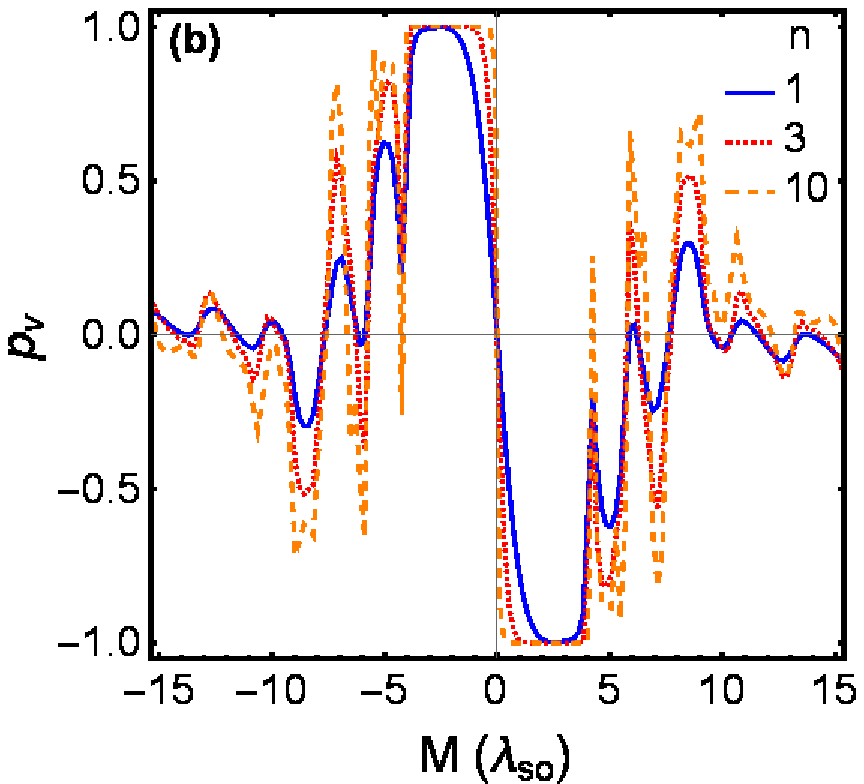}
\vspace*{-0.3cm}
\caption{(Colour online) Spin (a) and valley (b) polarizations through $n$ barriers versus exchange field $M$. 
Parameters used: $d=100$ nm, $w=50$ nm, $E_z=5 E_{c}$ , $E=40$ meV and $U=50$ meV}
\label{Fig:SpinValleyPolarization}
\end{figure}

Because evanescent tunneling is suppressed with the number of barriers, the spin $p_s$ and valley $p_v$ polarizations are altered as shown in Fig. \ref{Fig:SpinValleyPolarization} versus $M$. For one barrier this is similar to the $p_s$ or $p_v$ results (versus $U/E_F$) of Ref. \cite{VV}. Notice, though, that the $M$ ranges of near perfect $p_s$ and $p_v$ widen with the number of barriers.
	
\begin{figure}[t]
\hspace*{-0.4cm}
\includegraphics[
width=3.8cm,height=4cm
]{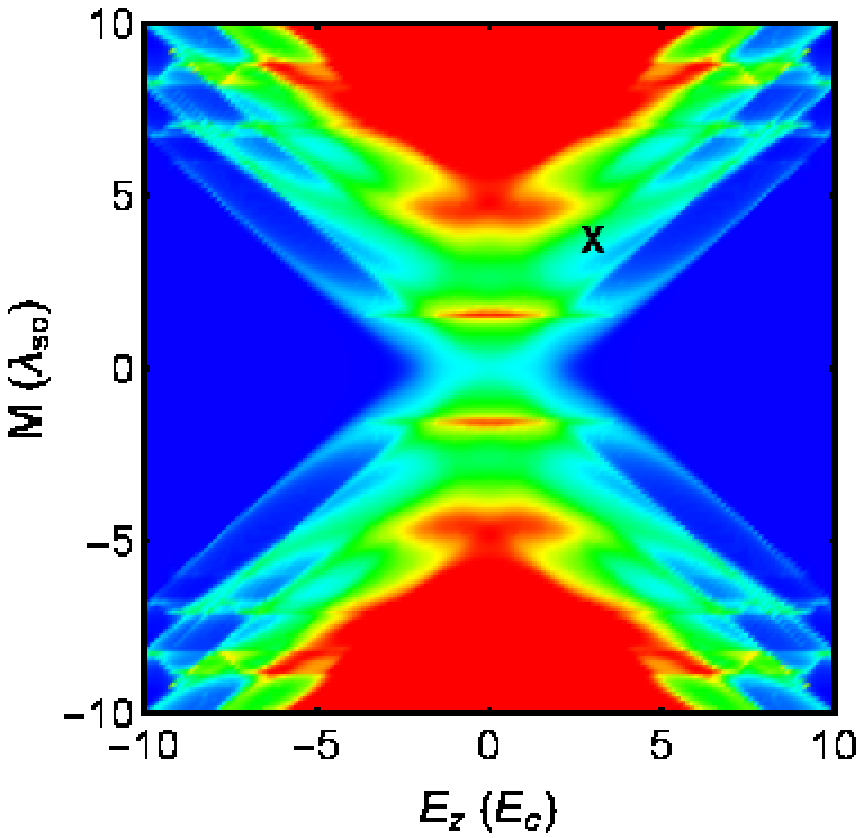}
\includegraphics[
width=3.8cm,height=4cm
]{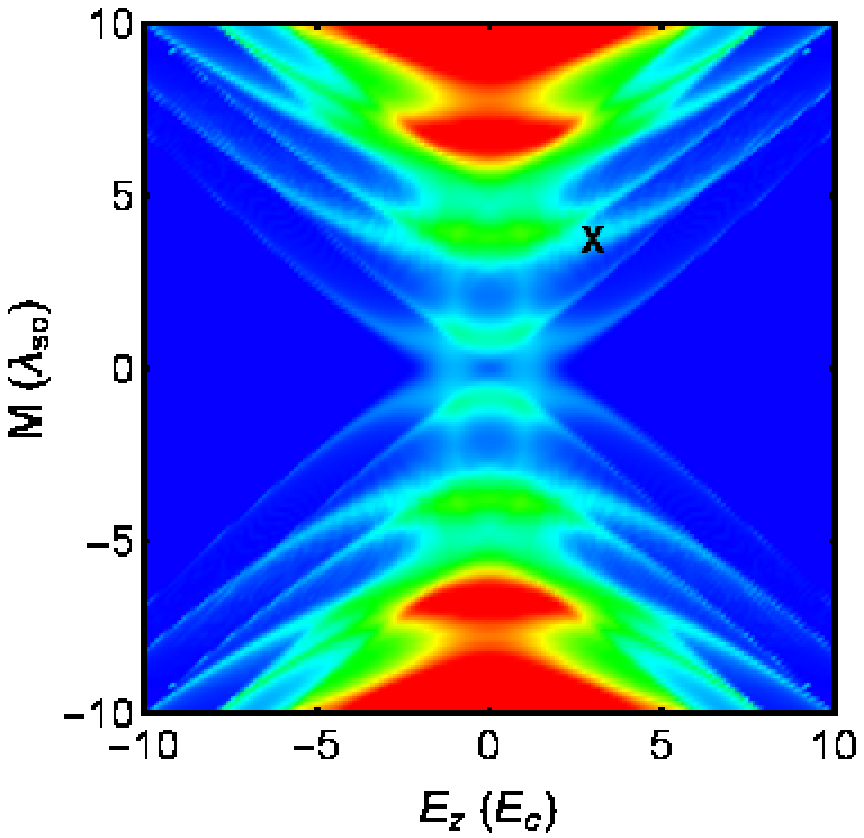}
\includegraphics[
width=0.6cm,height=3.8cm
]{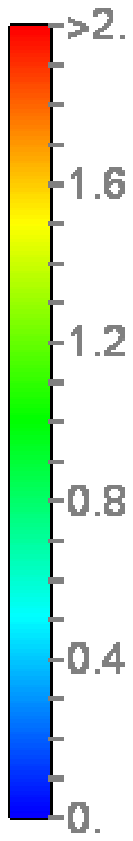}
\includegraphics[
width=3.8cm,height=4cm
]{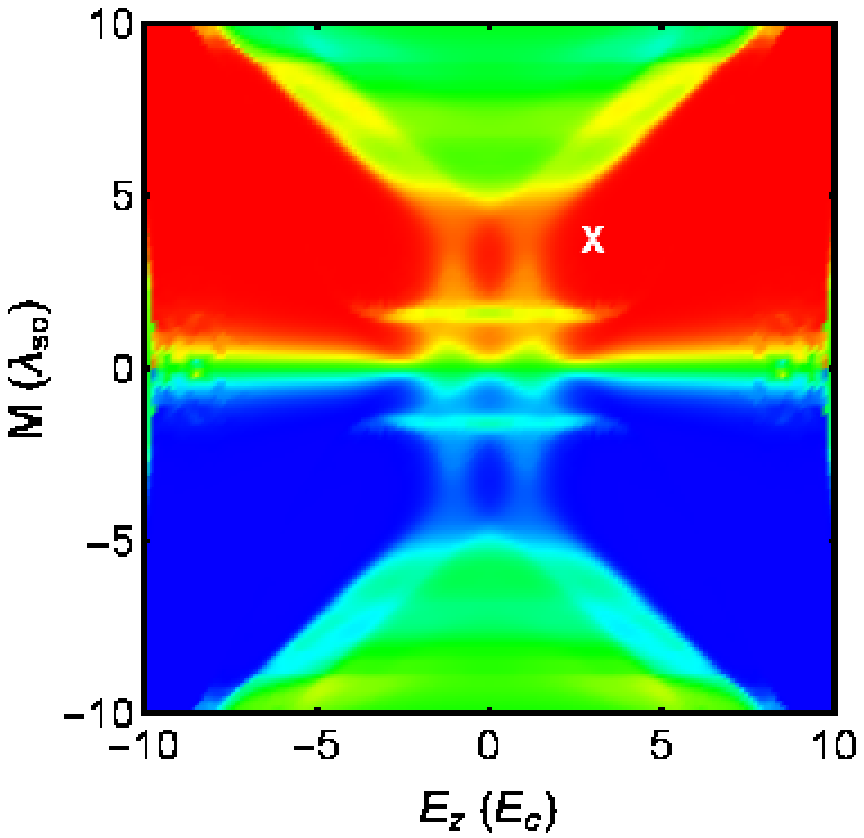}
\includegraphics[
width=3.8cm,height=4cm
]{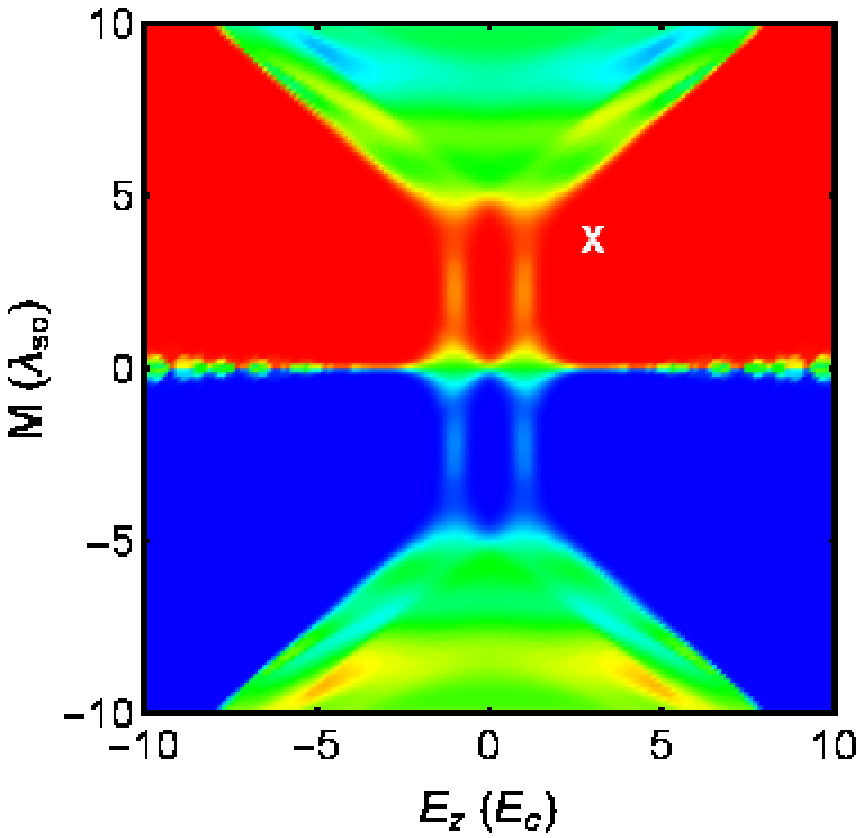}
\includegraphics[
width=0.79cm,height=3.85cm
]{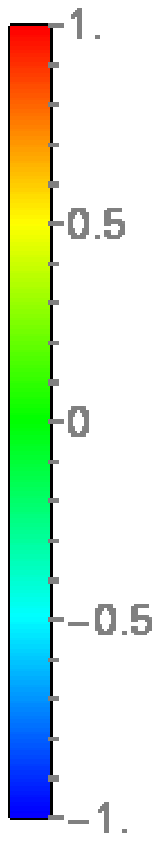}
\includegraphics[
width=3.8cm,height=4cm
]{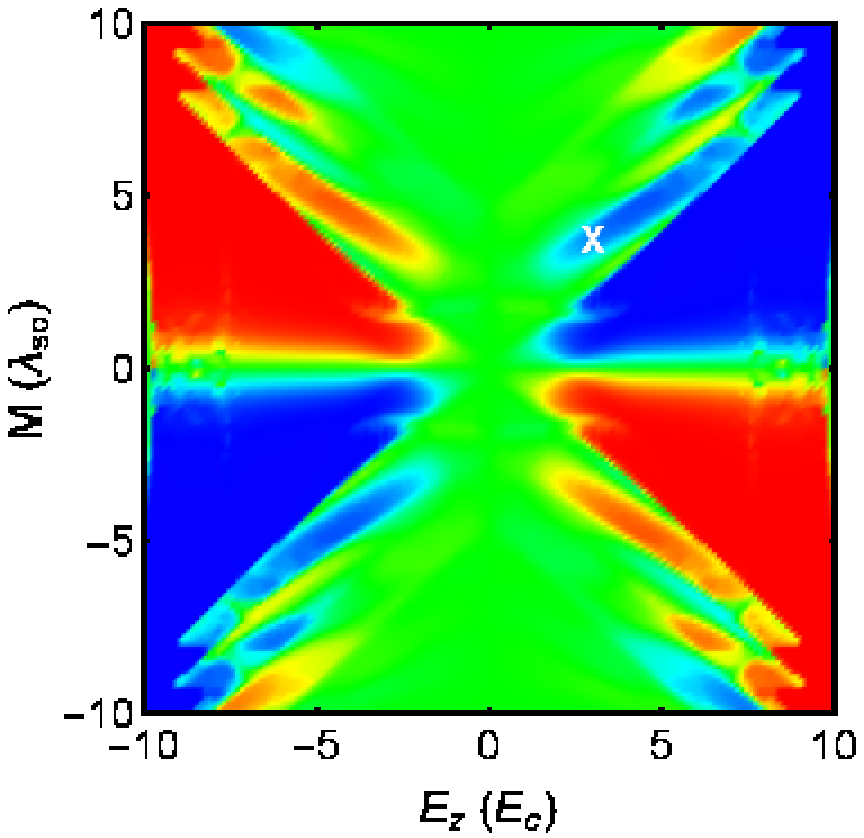}
\includegraphics[
width=3.8cm,height=4cm
]{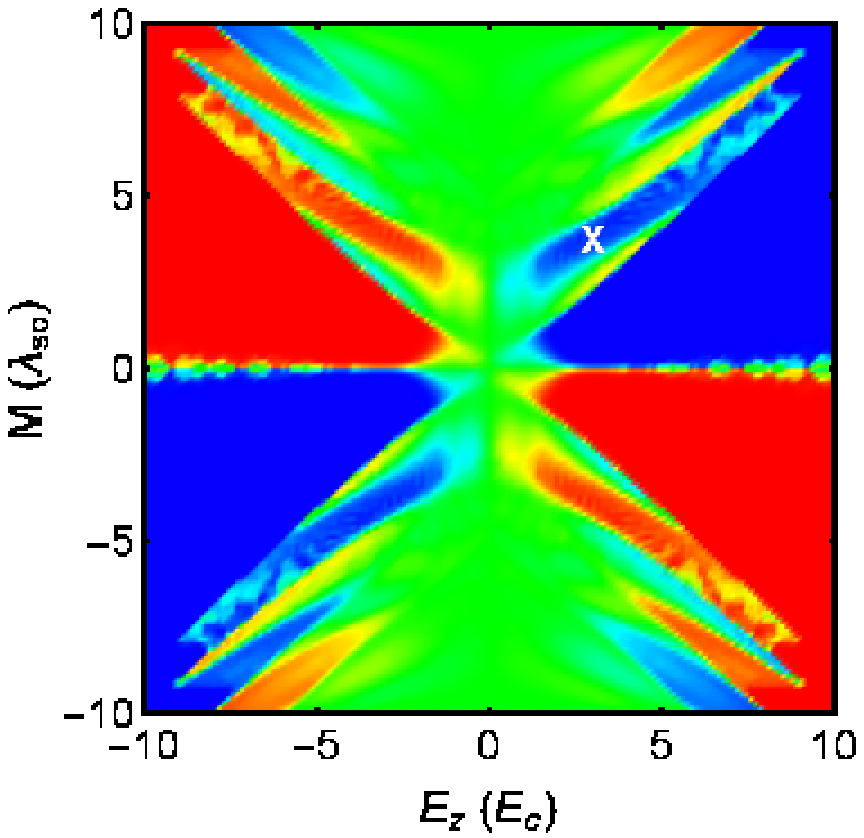}
\includegraphics[
width=0.79cm,height=3.85cm
]{LegendSpinVal.eps}
\vspace*{-0.6cm}
\caption{ (Colour online) Conductance (top), spin (middle) and valley polarization (bottom) through $2$ (left column) and $10$ (right column) barriers as a function of the electric and exchange field. The unit for the conductance plots is $g_0$. The symbol $\times$ indicates the parameters for which the current is polarized in the $K^{\prime}$-valley and only consists of spin-up particles. Parameters used: $d=100$ nm, $w=50$ nm, $E=40$ meV and $U=50$ meV. }
\label{Fig:CompareConductancePolarization}
\end{figure}

In Fig. \ref{Fig:CompareConductancePolarization} we show the dependence of the conductance, spin and valley polarization on the fields $E_z$ and  $M$.  
The results show that it is possible to achieve independently a strong spin and valley polarization by a proper tuning of these fields. 
For example, at the position marked with $\times$, the current is polarized in the $K^{\prime}$ valley ($p_{v} = -1$) but also  consists only of spin-up particles ($p_{s}=1$). The use of multiple barriers makes this polarization more pronounced. The maps in Fig. \ref{Fig:CompareConductancePolarization} enable one to select the desired $p_{s}$ and $p_{v}$ 
by tuning the fields $E_z$ and  $M$. 

\begin{figure}[t]
\hspace*{-0.4cm}
\includegraphics[
width=4cm,height=4.05cm
]{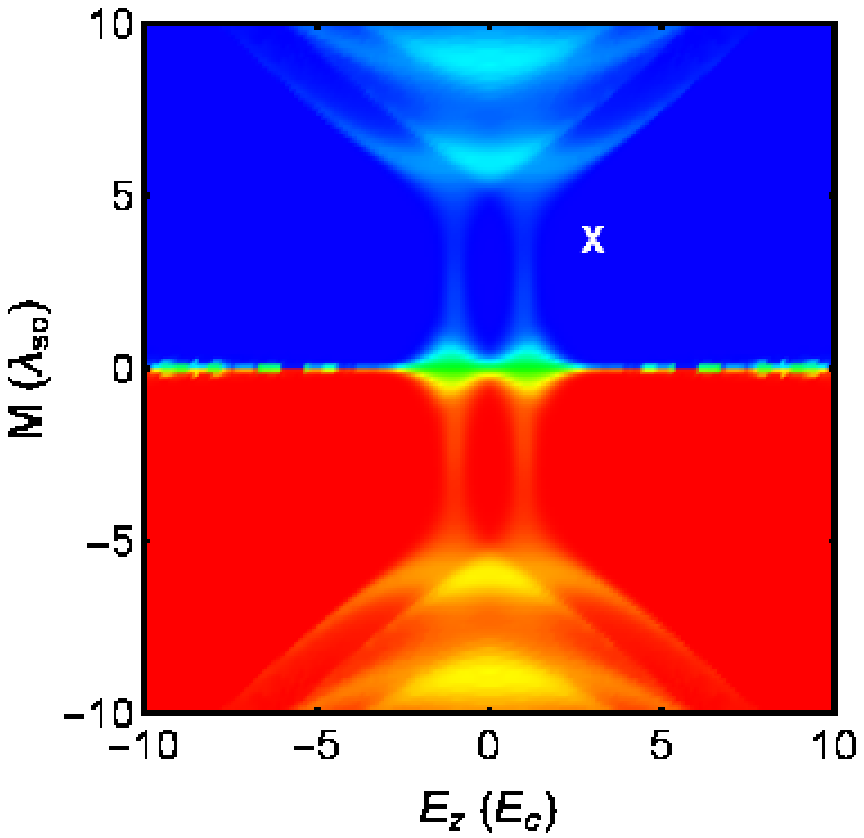}
\includegraphics[
width=3.85cm,
height=4.05cm
]{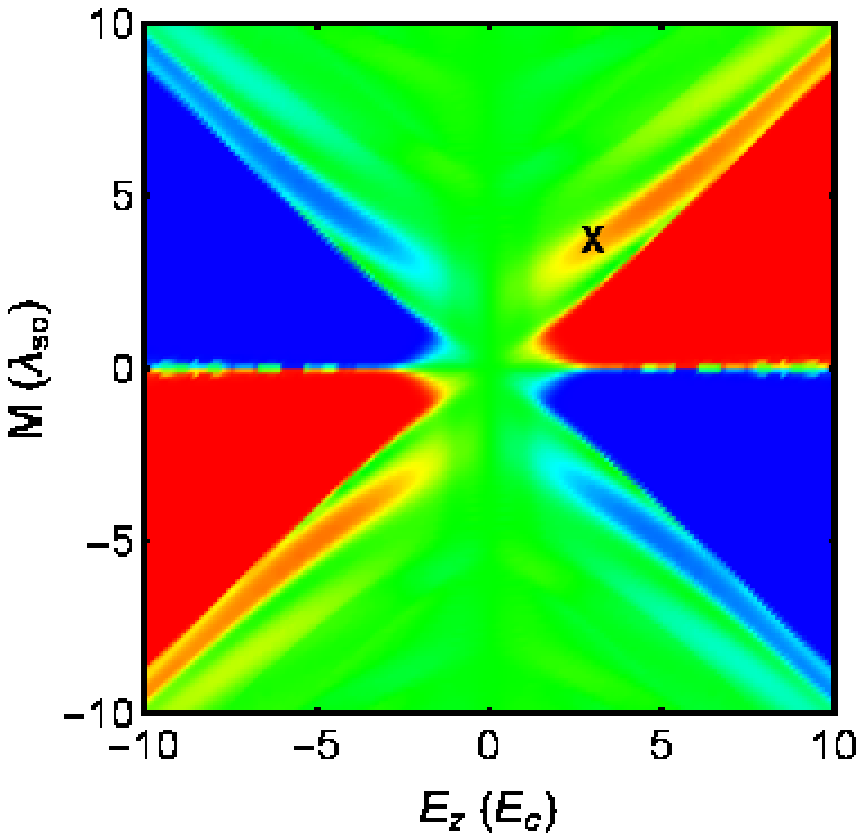}
\includegraphics[
width=0.65cm,height=3.92cm
]{Legend01.eps} 
\vspace*{-0.3cm}
\caption{(Colour online) ($M, E_z$) contour plots of the spin (left) and valley (right) polarizations through 10 barriers %as a function of the electric and exchange field 
for $E_F>U$. The position of the symbol $\times$ is the same as in Fig. \ref{Fig:CompareConductancePolarization}. Parameters used:  $d=100$ nm, $w=50$ nm, $E=60$ meV, and $U=50$ meV. }
\label{Fig:HighEPolarization}
\end{figure}

The spin and valley polarizations in Fig. \ref{Fig:CompareConductancePolarization} are shown for the Fermi level just beneath the height of the potential barrier $U$, i.e., $E_F\equiv E = 0.8 U$. In contrast, in Fig. \ref{Fig:HighEPolarization}  we shown them 
for 10 barriers in the reverse case, i.e., for  $E_F=E=1.2 U$. The results are opposite to the ones shown in Fig. \ref{Fig:CompareConductancePolarization} and show that tuning $E_F$ 
is another way of selecting the desired spin and valley polarization. Again considering the position of the $\times$ symbol,  the current is $K$-valley-   ($p_v = 1$) and spin-down polarized ($p_s=-1$).


%%%%%%%%%%%%%%%%%%%%%%%%%%%%%%%%%%%%%%%%%%%%%%%%%%%%%%%%%%%%%%%%%
\begin{figure}[t]
\hspace*{-0.4cm}
\includegraphics[
width=2.61cm,height=4cm
]{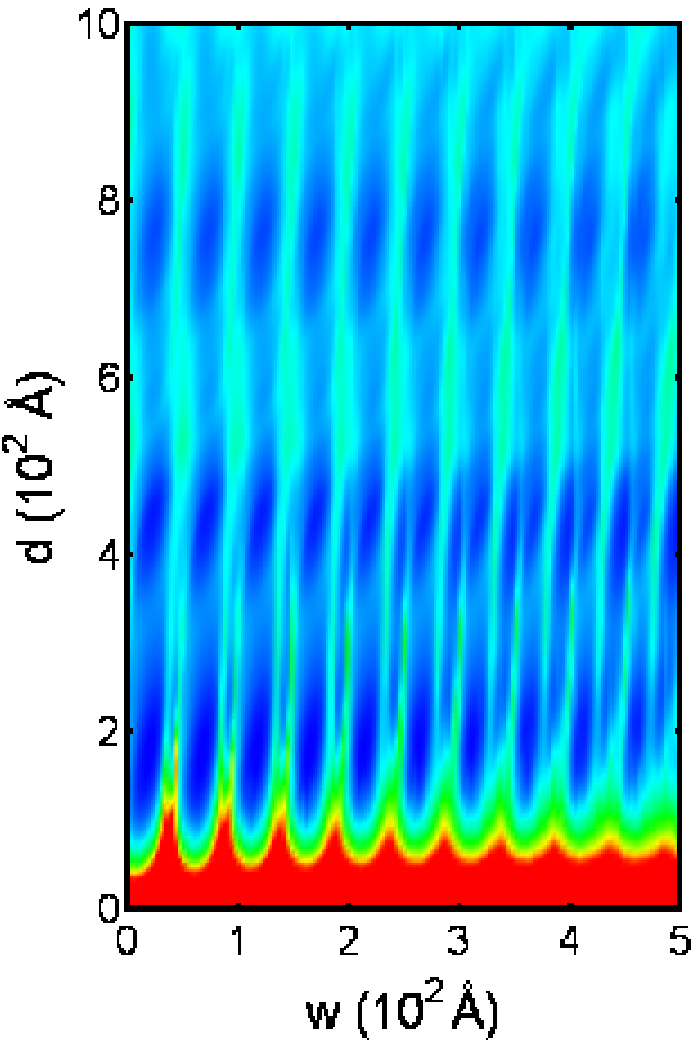}
\hspace*{-0.3cm}
\includegraphics[
width=0.54cm,height=3.92cm
]{LegendCond.eps}
\hspace*{-0.2cm}
\includegraphics[
width=2.61cm,height=4cm
]{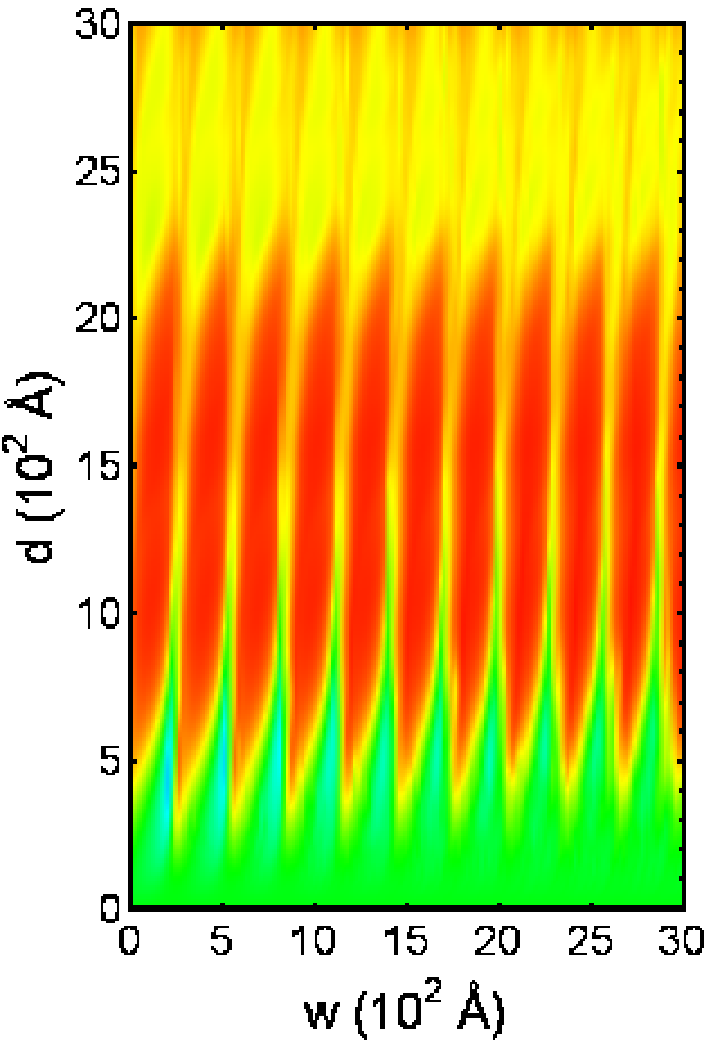}
\hspace*{-0.3cm}
\includegraphics[
width=2.61cm,height=4cm
]{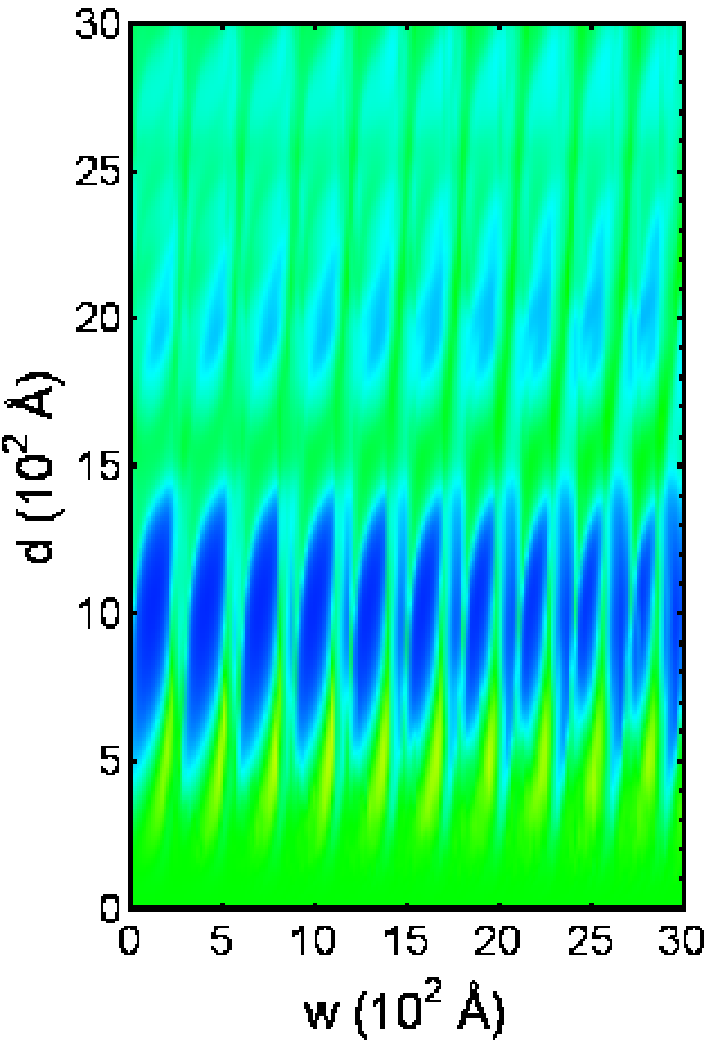}
\includegraphics[
width=0.65cm,height=3.92cm
]{LegendSpinVal.eps}
\vspace*{-0.6cm}
\caption{ (Colour online) ($d,w$) contour plot of the conductance (left) through $2$ barriers, of  the spin polarization
(center), and of the valley polarization (right) for $M=\lambda_{so}$ , $E_z=5E_{c}$, $E=40$ meV and $U=50$ meV. The unit for the conductance plots is $g_0$.}
\label{Fig:dwPlots}
\end{figure}

%%%%%%%%%%%%%%%%%%%%%%%%%%%%%%%%%%%%%%%%%%%%%%%%%%%%%%%%%%%%%%%%%%%%%%%%
%
We can also exploit the additional degree of freedom given by the barrier separation $w$ to tune the polarization. In Fig. \ref{Fig:dwPlots} we show a $(d,w)$ contour plot of the conductance and polarizations for $two$ barriers. As  shown, increasing the barrier width $d$ has a progressively detrimental  effect on the conductance. But, thanks to the periodic dependence of all these quantities at appropriate values of $d$, it is possible to realize a pure spin polarization and a valley mixed state as shown in Fig. \ref{Fig:dwPlots}. 
	
\begin{figure}[t]
\hspace*{-0.4cm}
\includegraphics[
width=2.61cm,height=4cm
]{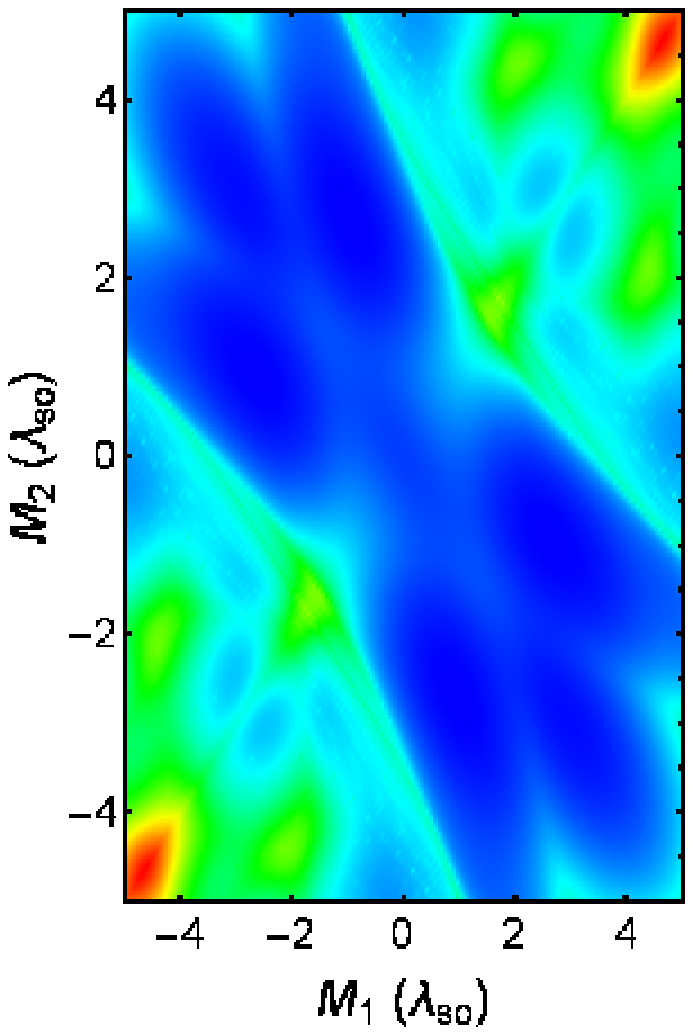}
\hspace*{-0.3cm}
\includegraphics[
width=0.54cm,height=3.92cm
]{LegendCond.eps}
\hspace*{-0.2cm}
\includegraphics[
width=2.61cm,height=4cm
]{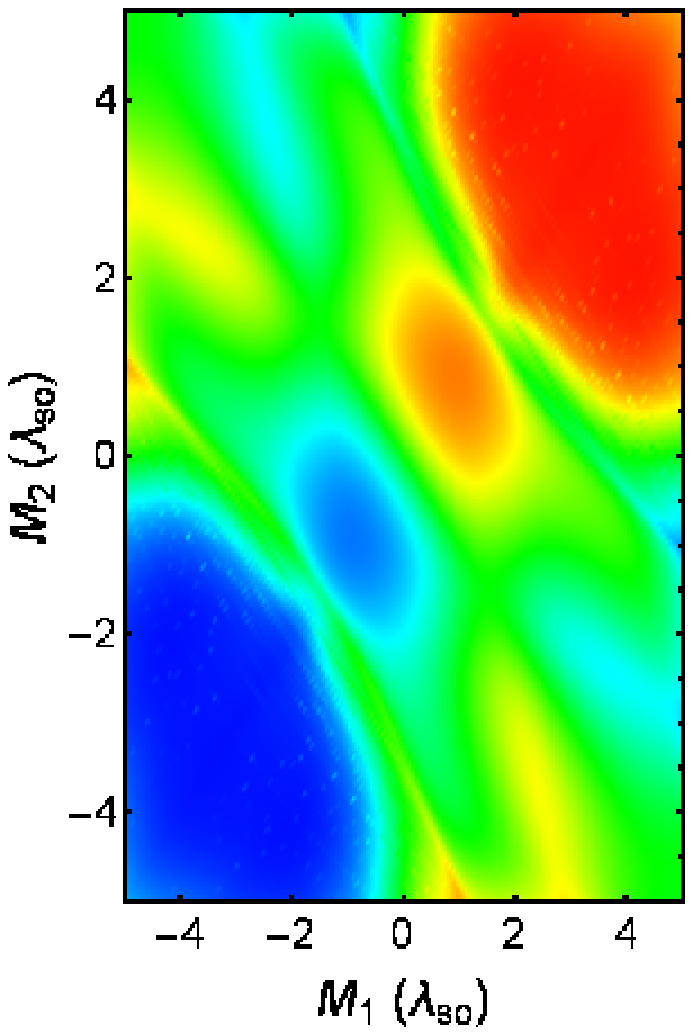}
\hspace*{-0.3cm}
\includegraphics[
width=2.61cm,height=4cm
]{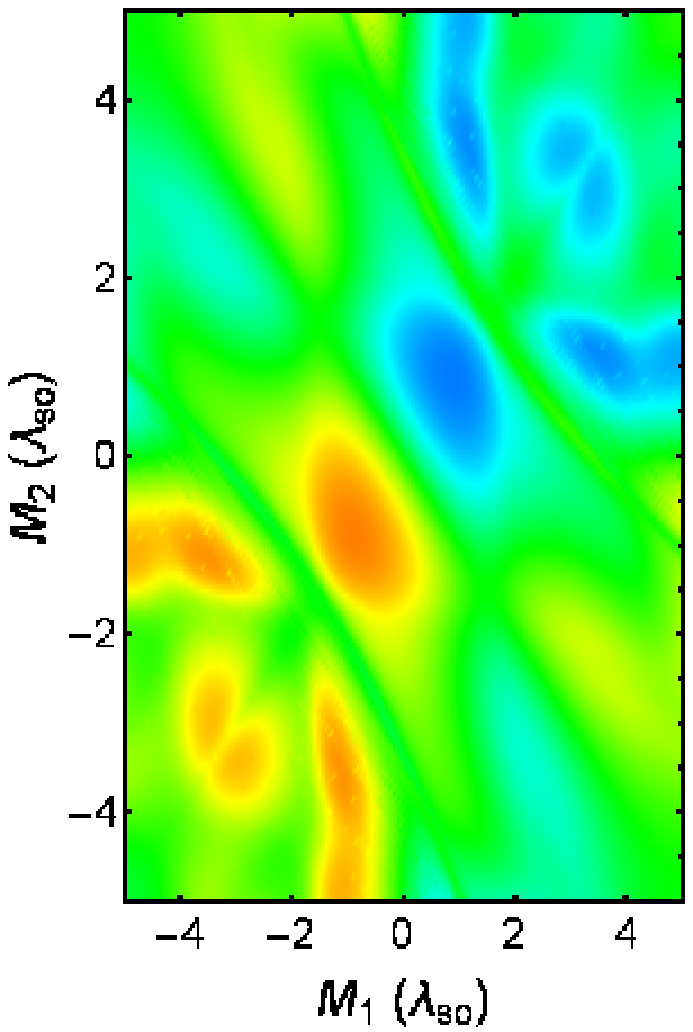}
\includegraphics[
width=0.65cm,height=3.92cm
]{LegendSpinVal.eps}
\vspace*{-0.6cm}
\caption{(Colour online) $(M_1, M_2)$  contour plot of the conductance (left) through $2$ barriers, of  the spin polarization
(center), and of the valley polarization (right). Parameters used:  $E_z=2E_{c}$, $d_1=d_2=100$ nm, $w=50$ nm, $E=40$ meV and $U=50$ meV. The unit for the conductance plots is $g_0$. }
\label{Fig:M1M2Plot}
\end{figure}

\begin{figure}[t]
\hspace*{-0.4cm}
\includegraphics[
width=2.61cm,height=4cm
]{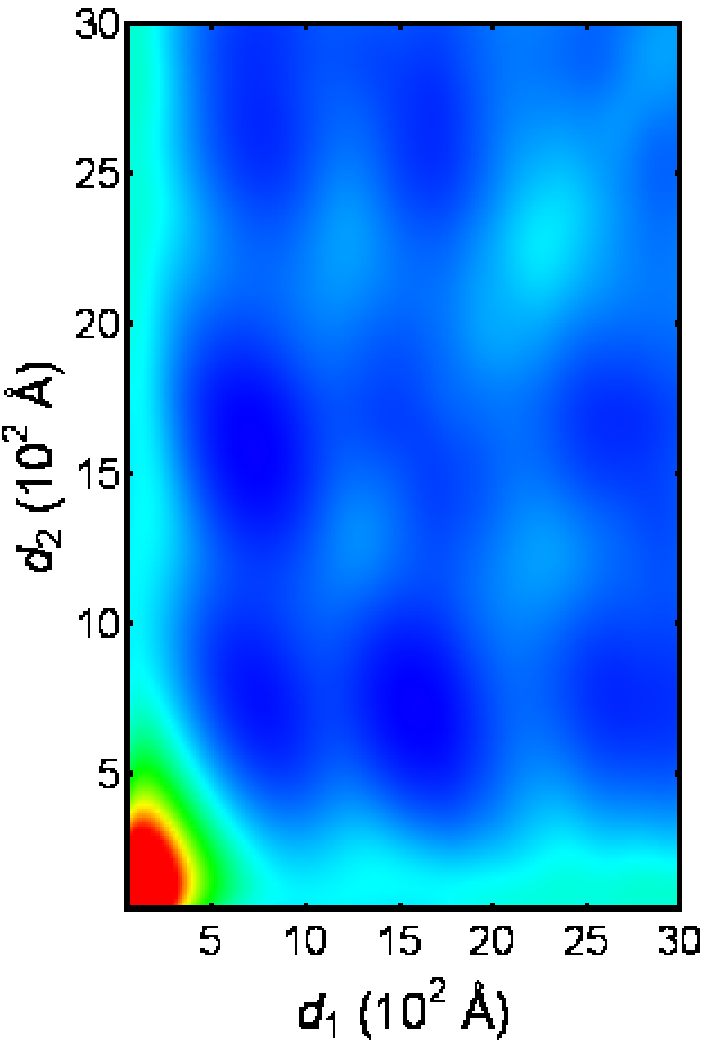}
\hspace*{-0.3cm}
\includegraphics[
width=0.54cm,height=3.9cm
]{LegendCond.eps}
\hspace*{-0.2cm}
\includegraphics[
width=2.61cm,height=4cm
]{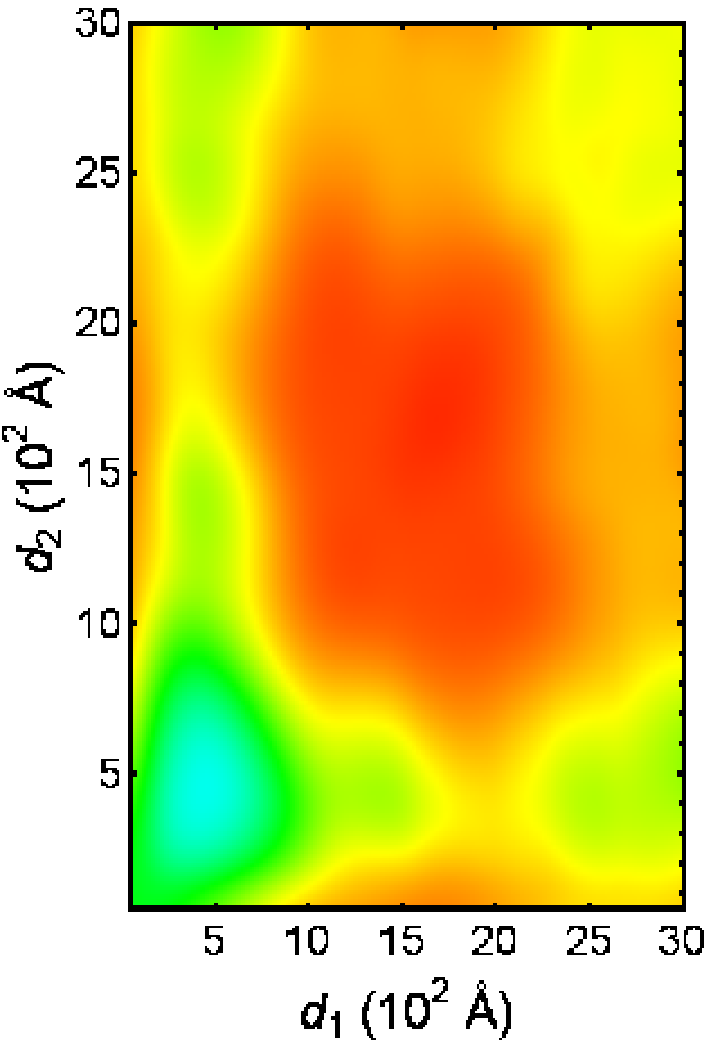}
\hspace*{-0.3cm}
\includegraphics[
width=2.61cm,height=4cm
]{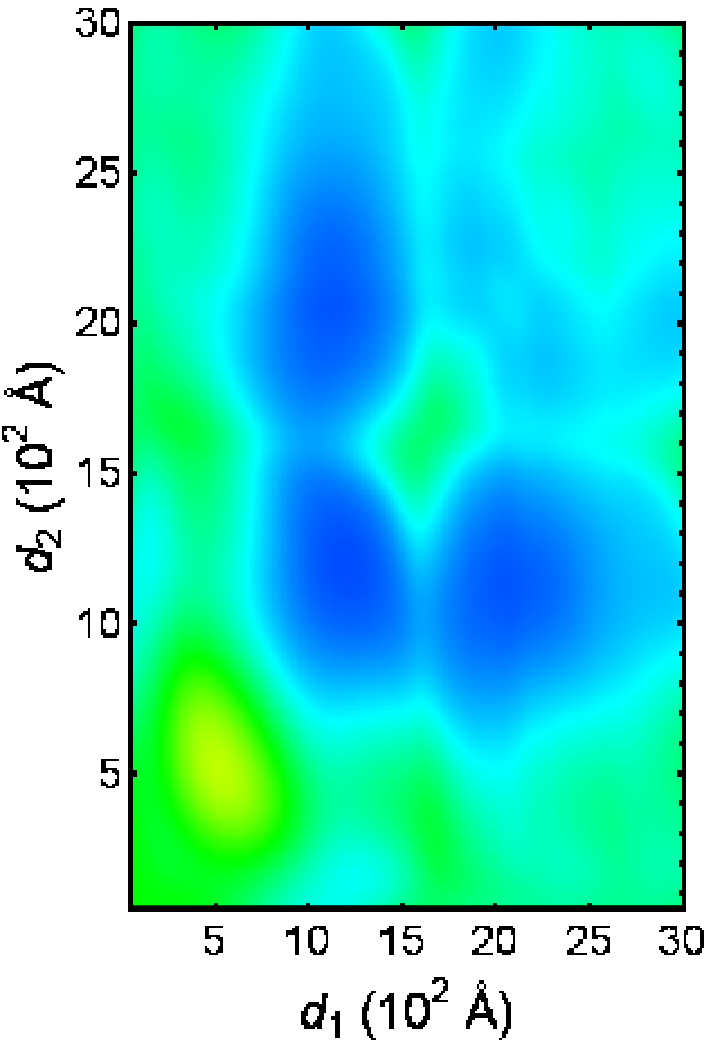}
\includegraphics[
width=0.65cm,height=3.9cm
]{LegendSpinVal.eps}
\vspace*{-0.5cm}
\caption{(Colour online) $(d_1, d_2)$  contour plot of the conductance (left) through $2$ barriers, of  the spin polarization
(center), and of the valley polarization (right). Parameters used: $E_z=2E_{c}$, $M_1=M_2=\lambda_{so}$, $w=50$ nm, $E=40$ meV and $U=50$ meV. The unit for the conductance plots is $g_0$.}
\label{Fig:d1d2Plot}
\end{figure}

We further explore the spin and valley polarizations by changing the magnetization $M$ of the two barriers independently at fixed barrier width $d$, as shown in Fig. \ref{Fig:M1M2Plot}, and by changing the width at fixed $M$, as shown in Fig. \ref{Fig:d1d2Plot}. In Fig. \ref{Fig:M1M2Plot} 
$M_1$ and $M_2$ are the FM fields in the first and second barrier, respectively, and in  Fig. \ref{Fig:d1d2Plot} $d_1$  and $d_2$ the corresponding widths. In either figure the left panels are for the conductance, the central ones for the spin polarization, and the right panels for the valley polarization. The central panel  in Fig. \ref{Fig:d1d2Plot} shows a near perfect spin-up polarization for a very wide range of $d_1$ and  $d_2$. This can become spin-down polarization if we reverse $M$ or change $U$. In either case we have a perfect spin filter, say, for $d_i\geq 7$ nm, $i=1,2$.


\subsection{Wells}

\begin{figure}[t]
\hspace*{-0.4cm}
\includegraphics[
width=3.95cm,height=4cm
]{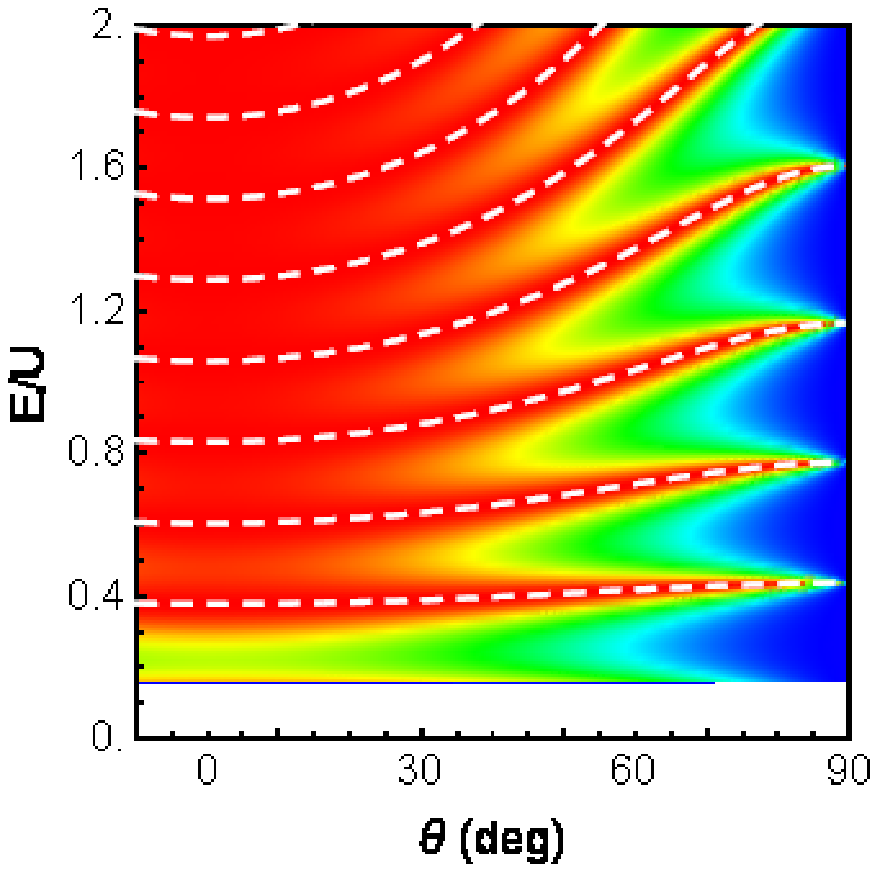}
\includegraphics[
width=3.75cm,height=4cm
]{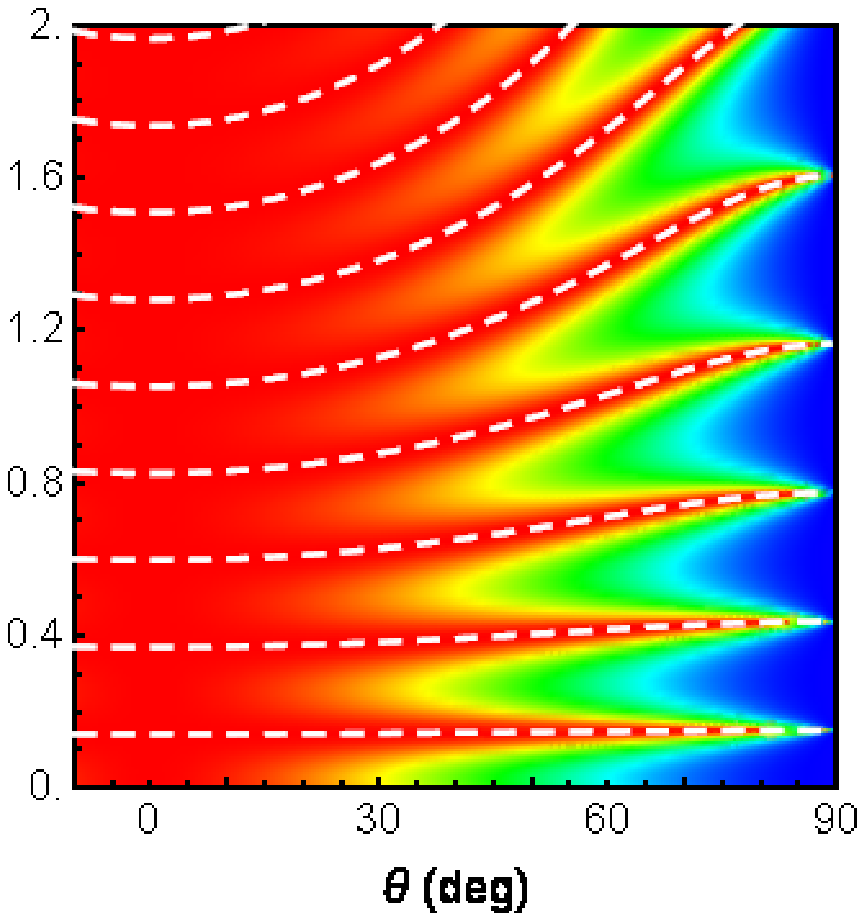}
\includegraphics[
width=0.8cm, height=4cm
]{Legend01.eps} 
\hspace*{-0.4cm}
\includegraphics[
width=3.95cm, height=4cm
]{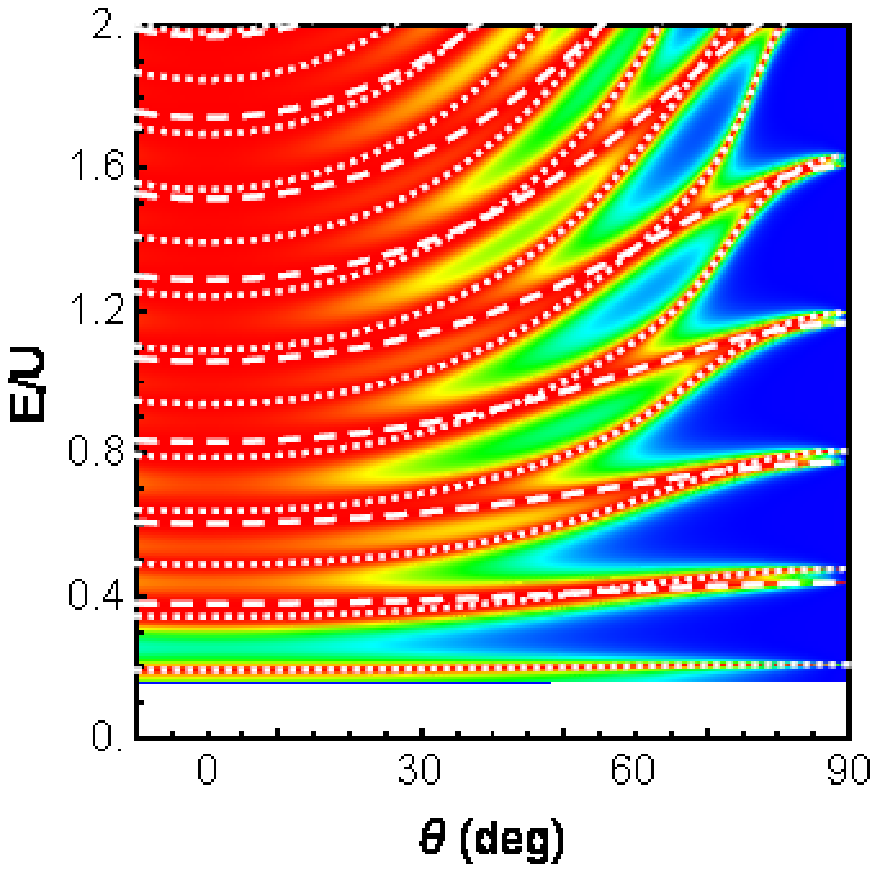}
\includegraphics[
width=3.75cm,height=4cm
]{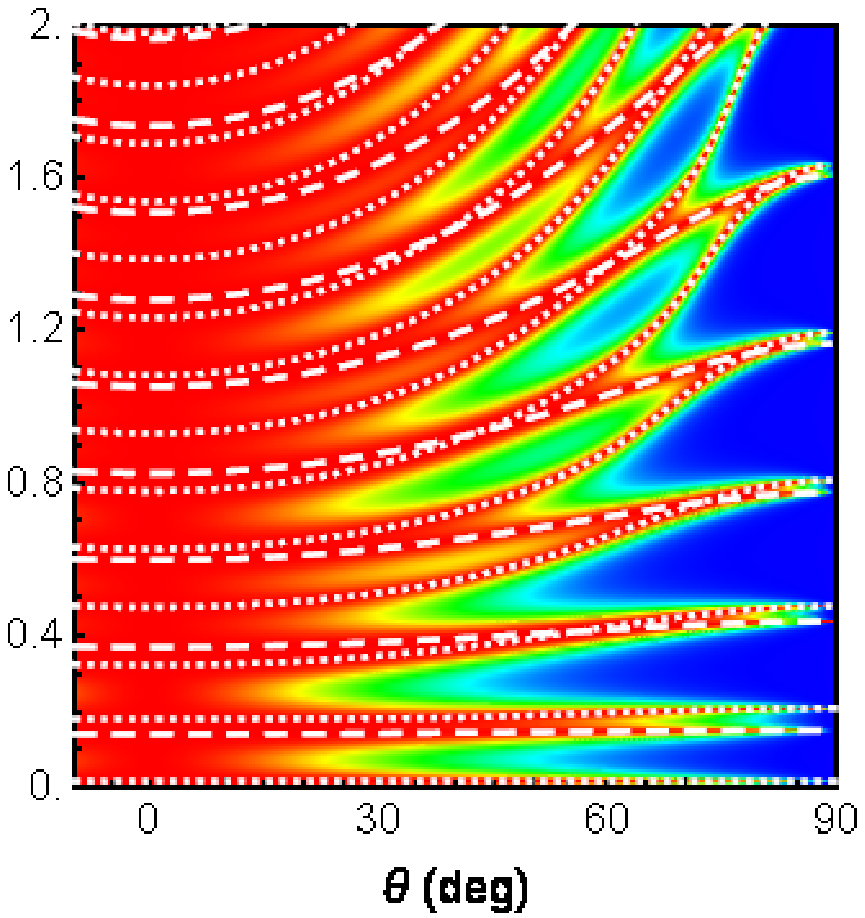}
\includegraphics[
width=0.8cm, height=4cm
]{Legend01.eps} 
\hspace*{-0.4cm}
\includegraphics[
width=3.95cm, height=4cm
]{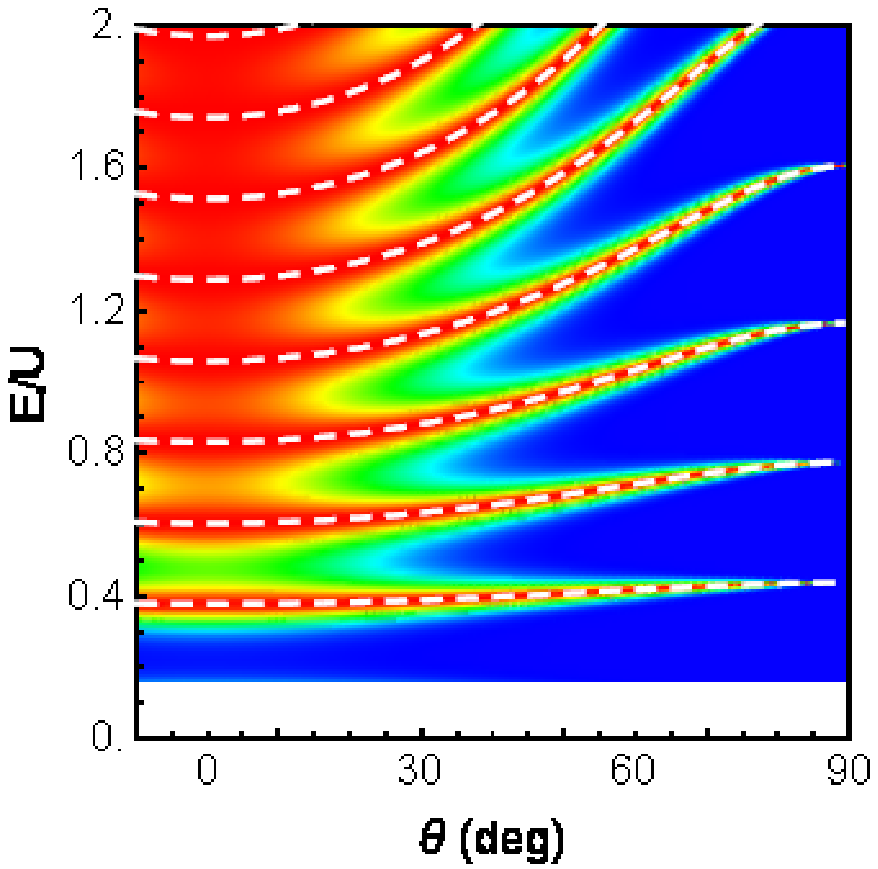}
\includegraphics[
width=3.75cm, height=4cm
]{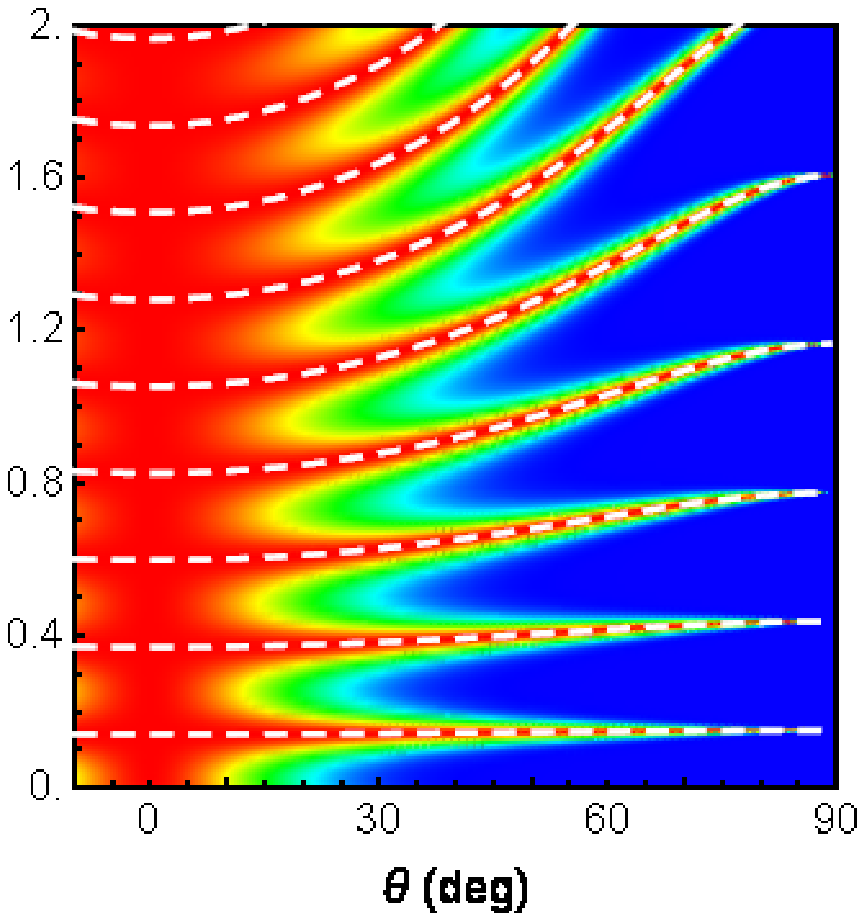}
\includegraphics[
width=0.8cm, height=4cm
]{Legend01.eps} 
\vspace*{-0.2cm}
\caption{(Colour online) $(E,\theta)$ contour plots of the transmission through one, two, and ten wells in the 1st, 2nd, and 3rd row respectively. The left (right) column is for spin-up (spin-down) electrons. The dashed white curves show the resonances calculated by Eq. (\ref{Eq:Resonances}) and the dotted white curves correspond to the solutions of $R_{\epsilon} = 0$. In the white region in the left column the transmission is undefined due to the lack of propagating states outside the barrier.
Parameters used are: $d=100$ nm, $w=50$ nm, $M=0 \lambda_{so}$ meV, $E_z=E_c$ and $U=-50$ meV}
\label{Fig:WellTransmission}
\end{figure}

Formally changing $U$ to $-U$ in Eq. (\ref{Eq:Transmission}) allows us to consider a set of wells as presented in Fig. \ref{Fig:fig1}(e).  In  Fig. \ref{Fig:WellTransmission} we show $(E,\theta)$ contour plots of the transmission in the presence of $1, 2$, and $10$ wells. The left (right) column is for spin-up (spin-down) electrons. As seen, similar to the case of graphene \cite{milt}, for certain angles the transmission is periodic in energy. Similar to the case of barriers, for the double well there is a new resonance pattern appearing and for a large number of wells the resonances become sharper and only those related to the single well case survive. Notice that the transmission is nearly perfect in a wider range of angles $\theta$ compared to that for barriers. 

\begin{figure}[t]
\includegraphics[width=4.2cm,height=4cm
]{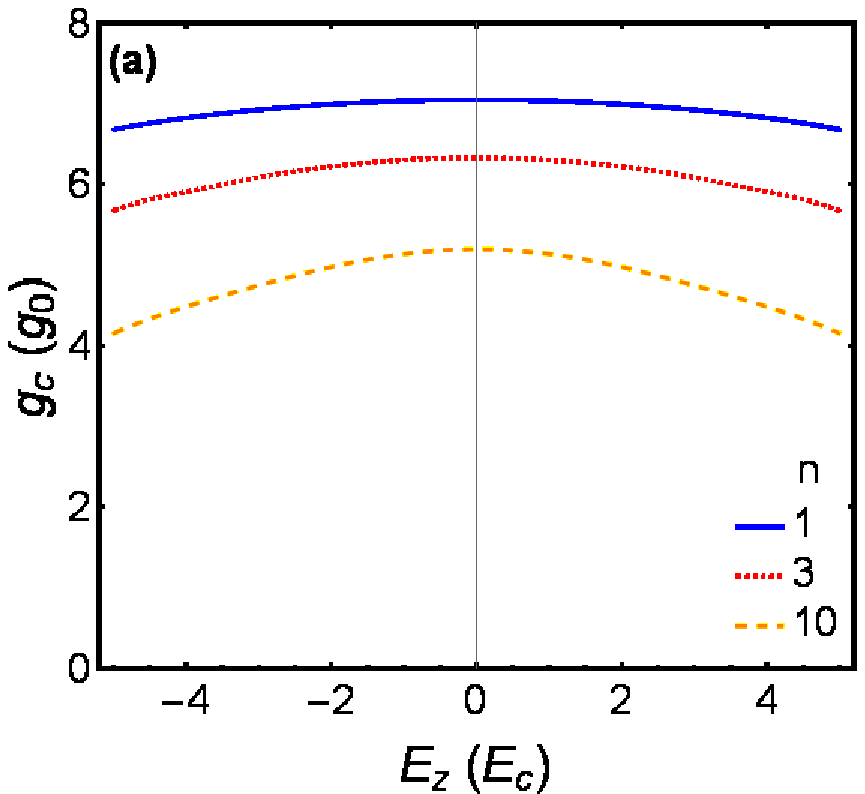} 
\includegraphics[width=4.2cm,height=4cm
]{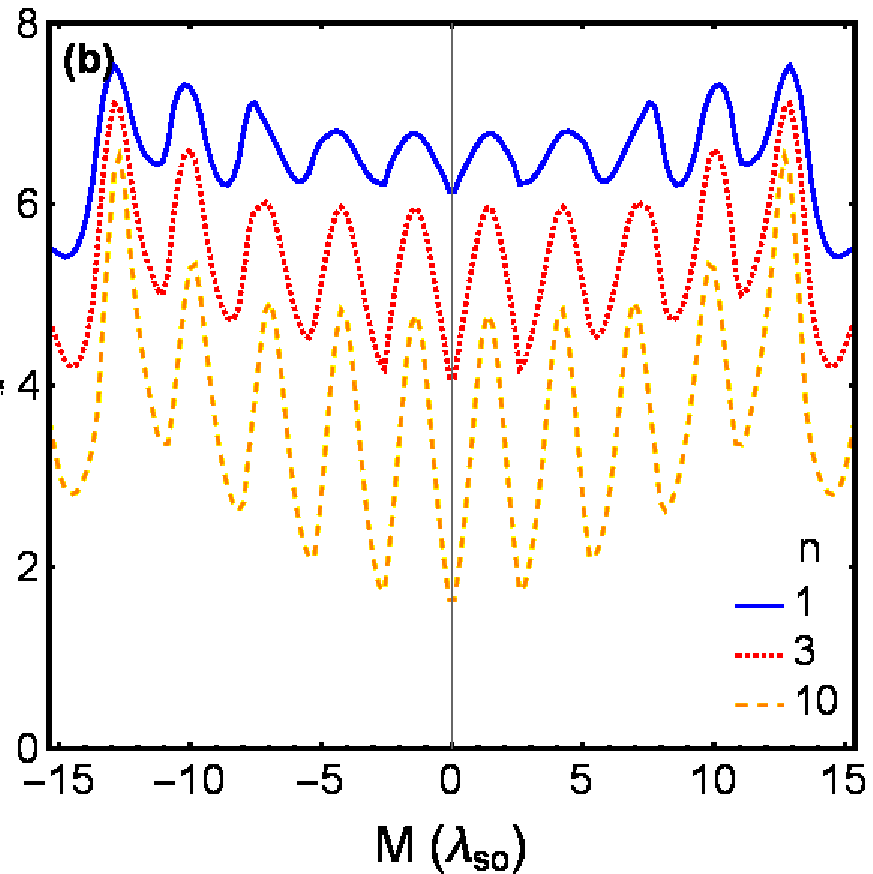}
\vspace*{-0.3cm}
\caption{(colour online) Conductance over  $n$ wells versus (a) the electric field for $M =\lambda_{so}$ and (b) the exchange field for $E_z = 5 E_{c}$.
The other parameters are $d=100$ nm, $w=50$ nm, and $U=-50$ meV.}
\label{Fig:WellConductance}
\end{figure}

As far as the conductance is concerned, Fig. \ref{Fig:WellConductance} shows that its overall behaviour, versus $E_z$ and $M$, is similar to that for barriers shown in Fig. \ref{Fig:ConductanceFirst}. However, we don't have a gap versus $M$, as in Fig. \ref{Fig:ConductanceFirst} because now the modes are always propagating. As for the spin  and  valley  polarizations, they are about one order of magnitude smaller than those involving barriers and are not shown.
	
\begin{figure}[t]
\includegraphics[width=4.2cm,height=4.0cm
]{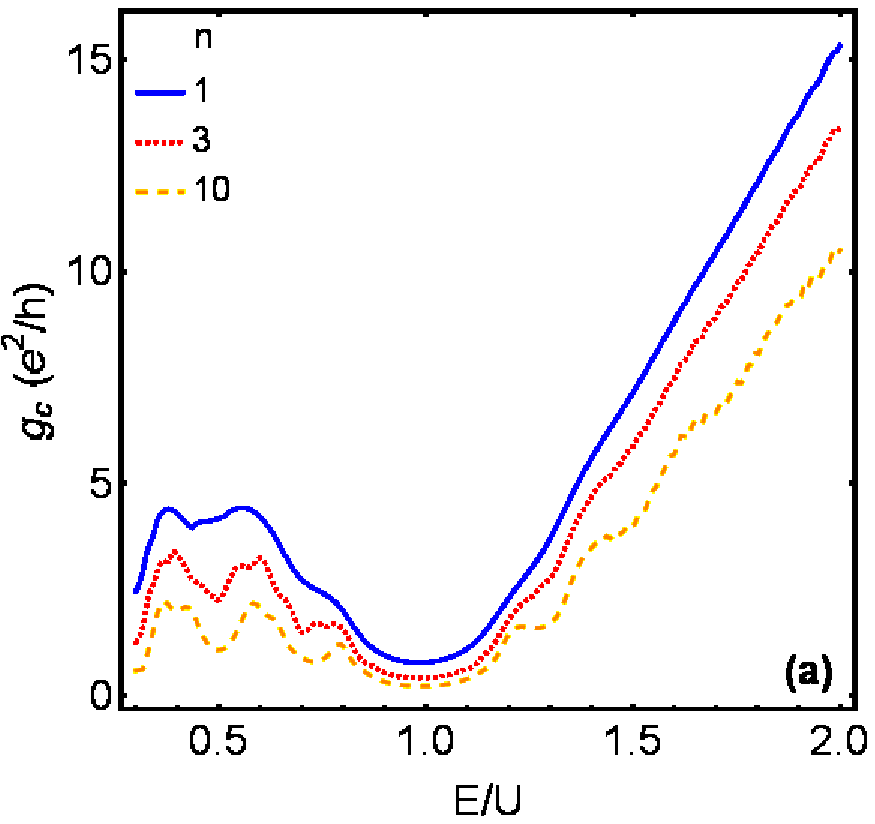}
\includegraphics[width=4.0cm,height=4.0cm
]{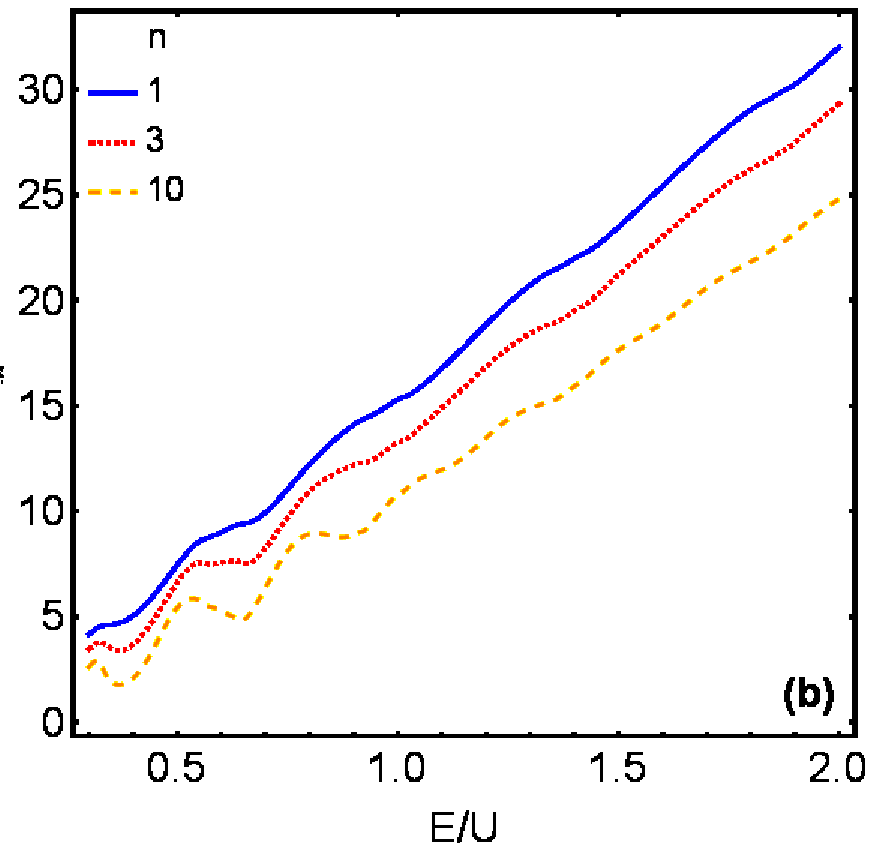}
\vspace*{-0.3cm}
\caption{(Colour online) Conductance through $n$ (a) barriers and (b) wells over a length $L_y=100$ nm as a function of the ratio $E/U$, ($E\equiv E_F$). 
The parameters are $d=100$ nm, $w=50$ nm, $M= \lambda_{so}$, $E_z= E_c$   $|U|=50$ meV.}
\label{Fig:WellBarrierConductance}
\end{figure}

As a function of the energy, the conductance is shown in Fig. \ref{Fig:WellBarrierConductance} for $n$ barriers on the left and for $n$ wells on the right. Notice the difference in the vertical scales and the overall reduction with increasing $n$. For $n=1$ the behaviour is similar to that of Refs. \cite{TY, VV} for silicene and Ref. \cite{milt} for graphene. The reduction with increasing $n$ can be understood by the fact the conductance is mainly governed by evanescent tunneling of modes that are near resonant. This results in a smearing of the resonances and 
of the transmission due to Klein-tunneling and, consequently, a  decrease of the  conductance in the  multi-barrier or multi-well system.

\vspace*{-0.6cm}
\section{Summary and conclusions}\label{Sec:Summary}
The use of multiple barrier and well structures allows for a new approach in  searching for  tunable valley and spin polarizations in silicene. Our analytical results help to clearly comprehend  the transmission resonances in  multi-barrier and multi-well structures. 

We found that a transport gap in the conductance develops not only when $U$ is varied \cite{TY,VV} but also when $M$ is.
For multiple barriers this gap, as well as the spin and valley polarizations,   widen with  $n$ because of the suppression of nearly propagating modes. This same mechanism also sharpens the resonances found in single barriers. 

The quantitative assessment of the conductance and of the polarizations, as functions of the applied electric and exchange fields, the width of the barriers or wells and their separation, suggests a selection of parameters to use in order to obtain the desired spin and valley polarizations. 
In this respect, one may wonder how reasonable the parameters we used are. As a matter of fact, a typical 
$E_z$ value is \cite{jang}  $V/6$ nm and one for $M$, though for graphene \cite{brat}, is $M=3$ meV. Additionally, the use of electric field strengths up to 2.7 V/nm have been reported for bilayer graphene \cite{ExpEz}. Accordingly, the range of the $E_z$ and $M$ values used in our calculations as well as some critical values of theirs are entirely reasonable. 
We also notice that all these quantities oscillate nearly periodically with the separation between barriers or wells.

Finally we showed that for wells the conductance oscillates with the exchange field and that the transport gap observed for barriers is absent.

\vspace*{-0.6cm}
\section*{Acknowledgments}
This work was supported by the Canadian NSERC Grant No. OGP0121756 (PV) and by the Flemish Science Foundation (FWO-Vl) with a Ph. D. research grant (BVD).

\appendix
\vspace*{-0.4cm}
\section*{Calculation of the transmission}
The Hamiltonian for silicene with a perpendicular electric field $E_{z}$ and
magnetization M at the $K$ point in the basis $\{\phi _{A}^{\uparrow },\phi
_{A}^{\downarrow },\phi _{B}^{\uparrow },\phi _{B}^{\downarrow }\}$ is given by
\begin{equation}
H_{K}(\vec{k})=%
\begin{pmatrix}
E_1(1,1) & i\lambda _{R}a\hat{k}_{-} & \upsilon_{F}\hat{k}_{-} & 0 \\ 
-ia\lambda _{R}\hat{k}_{+} & E_1(-1,1) & 0 & \upsilon_{F}\hat{k}_{-} \\ 
\upsilon_{F}\hat{k}_{+} & 0 & E_1(1,-1) & -ia\lambda _{R}\hat{k}_{-} \\ 
0 & \upsilon_{F}\hat{k}_{+} & ia\lambda _{R}\hat{k}_{+} & E_1(-1,-1)
\end{pmatrix}%
\end{equation}%
and at the $K^{\prime}$ point by 
\begin{equation}
H_{K^{\prime }}(\vec{k})=%
\begin{pmatrix}
E_{-1}(1,1) 
& -i\lambda _{R}a\hat{k}_{-} & \upsilon_{F}\hat{k}_{+} & 0 \\ 
ia\lambda _{R}\hat{k}_{+} & E_{-1}(-1,1)  & 0 & \upsilon_{F}\hat{k}_{+} \\ 
\upsilon_{F}\hat{k}_{-} & 0 & E_{-1}(1,-1)  & ia\lambda _{R}\hat{k}_{-} \\ 
0 & \upsilon_{F}\hat{k}_{-} & -ia\lambda _{R}\hat{k}_{+} & E_{-1}(-1,-1) 
\end{pmatrix}%
\end{equation}%
where $a$ is the interatomic distance, $\lambda _{R}$ a very small \cite{ezawa1,ezawa3} SOI constant,  and
\begin{equation}
E_{\eta}(s_z,\tau_z)=\eta\lambda_{so} s_z\tau_z +lE_z\tau_z+M_zs_z.
\end{equation}

The wave functions $\Psi $ are given by the  matrix product 
\begin{equation}
\Psi _{j}=P_{j}E_{j}(x)C_{j}
\end{equation}%
which, for the $K$ point, are given by: 
\begin{eqnarray}
P_{1} &=&%
\begin{pmatrix}
1 & 1 & \mu  & \mu  \\ 
\zeta k_{1-}   & -\zeta  k_{1+}  & \eta  k_{2_-} & -\eta  k_{2+} \\ 
\xi k_{1+}  & -\xi k_{1-} & \chi k_{2+} & -\chi  k_{2-}  \\ 
\nu  & \nu  & 1 & 1%
\end{pmatrix}\,,
\\ \notag
\\
E_{1}(x) &=&%
\begin{pmatrix}
e^{ik_{1}x} & 0 & 0 & 0 \\ 
0 & e^{-ik_{1}x} & 0 & 0 \\ 
0 & 0 & e^{ik_{2}x} & 0 \\ 
0 & 0 & 0 & e^{-ik_{2}x}%
\end{pmatrix},
\end{eqnarray} %\notag \\
\begin{equation}
\hspace*{-0.7cm}C_{j} =
(A_{j}, \quad
B_{j}, \quad
C_{j}, \quad
D_{j})^T,
\end{equation}%
where $T$ denotes the transpose and  $k_{i\pm}=k_{i}\pm ik_{y},\,i=1,2$. Also, with $a_\lambda =a^{2}\lambda _{R}^2$ we have 
\begin{equation}
k_{1} =[k_{F1}^{2}-k_{y}^{2}]^{1/2},\quad
k_{2} =[k_{F2}^{2}-k_{y}^{2}]^{1/2},
\end{equation} 
\begin{equation}
k_{F1} =\big[ a_\lambda E_{+}^{2}+\upsilon_{F}^{2} E_{-}^{2} -\Delta _{2}"\big](a_\lambda+\upsilon_{F}^{2})^{-1/2},
\end{equation} 
\begin{equation} 
k_{F2} =\big[ a_\lambda E_{+}^{2}+\upsilon_{F}^{2}E_{+}^{2}+\Delta _{2}"\big](a_\lambda+\upsilon_{F}^{2})^{-1/2}, 
\end{equation} 
\begin{equation}
\Delta _{2}" =2\sigma\sqrt{a_\lambda^2 N^2+a_\lambda \upsilon_{F}^{2}\Omega +N^2 \upsilon_{F}^{4}}, 
\end{equation}
\begin{equation}
\sigma =sign[M(\Delta_zE+\lambda _{so}M)(a_\lambda \Delta_{z}+\upsilon_{F}^{2}M)],
\end{equation} 
\begin{equation}
\Omega = \Delta_{z}^{2}E_+^2+4\Delta_{z}E\lambda _{so}M+(E_+^2+\Delta_{z}^{2}) M^{2}-M^{4},
\end{equation}
where 
$E_\pm^2=E^{2}\pm \Delta_{z}^{2}-\lambda_{so}^{2}+M^{2}$ and $N^2=( \Delta_{z}E+\lambda_{so}M) ^{2}$. Further,
\begin{eqnarray}
\mu &=&\frac{ia\lambda _{R}\upsilon_{F} (\beta -\gamma) k_{F2}^{2} }
{\alpha\beta \gamma -(a^{2}\lambda _{R}^{2}\gamma+\beta \upsilon_{F}^{2})k_{F2}^{2} }, \\*
\nu  &=&\frac{-ia\lambda _{R}\upsilon_{F} (\beta -\gamma) k_{F1}^{2}} {\delta\beta \gamma- (a^{2}\lambda _{R}^{2}\beta+ \upsilon_{F}^{2} \gamma)k_{F1}^{2}}, \\*
 \notag 
 \\
\zeta  &=&\frac{ia\lambda _{R}+\upsilon_{F}\nu}{\beta },\,\, 
\eta  =\frac{ia\lambda _{R}\mu +\upsilon_{F}}{\beta},  \\ 
\xi  &=&\frac{ia\lambda _{R}\nu +\upsilon_{F}}{\gamma}, \,\,  
\chi  =\frac{ia\lambda _{R}+\upsilon_{F}\mu }{\gamma}.   
\end{eqnarray}%
\begin{eqnarray}
\alpha  &=&\left( E-lE_{z}-\lambda _{so}-M\right) \notag \\
\beta  &=&\left( E-lE_{z}+\lambda _{so}+M\right)   \notag \\
\gamma  &=&\left( E+lE_{z}+\lambda _{so}-M\right)   \notag \\
\delta  &=&\left( E+lE_{z}-\lambda _{so}+M\right)  
\end{eqnarray}%
and at the $K^{\prime}$ point:%
\begin{eqnarray}
P_{1}^{\prime } &=&%
\begin{pmatrix}
\eta ^{\prime }k_{1+}^{\prime }  & -\eta ^{\prime
}  k_{1-}^{\prime } & \zeta ^{\prime } 
k_{2+}^{\prime } & -\zeta ^{\prime }k_{2-}^{\prime
} \\ 
1 & 1 & \nu ^{\prime } & \nu ^{\prime } \\ 
\mu ^{\prime } & \mu ^{\prime } & 1 & 1 \\ 
\chi ^{\prime } k_{1-}^{\prime }  & -\chi ^{\prime
} k_{1+}^{\prime } & \xi ^{\prime } k_{2-}^{\prime
} & -\xi ^{\prime } k_{2+}^{\prime } 
\end{pmatrix}
\end{eqnarray}%
where $k_{i\pm}^\prime=k_{i}^\prime\pm ik_{y},\,i=1,2$. The matrix $E_{1}^{\prime }(x)$ is obtained
from $E_{1}(x)$ by changing $k_{i}$ to $k_{i}^\prime$. Also,  $k_{i}^\prime$ and $k_{Fi}^\prime$ are 
obtained from $k_{i}$ and $k_{Fi}$ by the same change followed by $\Delta" _{2}\to \Delta _{2}"^{\prime}$.

\begin{eqnarray}
\mu ^{\prime } &=&\frac{ia\lambda _{R}\upsilon_{F}(\alpha^{\prime }-\delta ^{\prime}) k_{F2}^{ 2}}
{\beta ^{\prime }\alpha ^{\prime }\delta^{\prime}- (a^{2}\lambda _{R}^{2} \delta^{\prime}+\upsilon_{F}^{2}\alpha^{\prime }) k_{F2}^{2}}
 \\
\nu ^{\prime } &=&\frac{-ia\lambda _{R}\upsilon_{F}(\alpha ^{\prime }-\delta ^{\prime})
 k_{F1}^{2}}
{\gamma^{\prime}\alpha ^{\prime }\delta^{\prime}-(a^{2}\lambda _{R}^{2}\alpha^{\prime }+\upsilon_{F}^{2}\delta ^{\prime })k_{F1}^{2}}\\ 
\notag
\\
\zeta^{\prime}  &=&\frac{ia\lambda _{R}+\upsilon_{F}\nu^{\prime}}{\alpha^{\prime}},\,\,
\eta^{\prime}  =\frac{ia\lambda _{R}\mu^{\prime} +\upsilon_{F}}{\alpha^{\prime}},  \\ 
\xi^{\prime}  &=&\frac{ia\lambda _{R}\nu^\prime +\upsilon_{F}}{\delta^\prime}, \,\, 
 \chi^{\prime}  =\frac{ia\lambda _{R}+\upsilon_{F}\mu^{\prime}}
 {\delta^{\prime}}.
\end{eqnarray}
\begin{eqnarray}
\alpha  &=&\left( E-lE_{z}-\lambda _{so}-M\right) \notag \\
\beta  &=&\left( E-lE_{z}+\lambda _{so}+M\right)   \notag \\
\gamma  &=&\left( E+lE_{z}+\lambda _{so}-M\right)   \notag \\
\delta  &=&\left( E+lE_{z}-\lambda _{so}+M\right)  
\end{eqnarray}
\begin{eqnarray}
\notag
\alpha ^{\prime } &=&\left( E-lE_{z}+\lambda _{so}-M\right)  \\
\beta ^{\prime } &=&\left( E-lE_{z}-\lambda _{so}+M\right)   \notag \\
\gamma ^{\prime } &=&\left( E+lE_{z}-\lambda _{so}-M\right)   \notag \\
\delta ^{\prime } &=&\left( E+lE_{z}+\lambda _{so}+M\right) 
\end{eqnarray}

To calculate the transmission through an arbitrary number of barriers n, we exploit the following property of the transfer matrix through {\it a single barrier} at $x=x_{0}$: 
\begin{eqnarray}
\notag
&&\hspace*{-7.7cm}M(x_{0})=E_{1}^{-1}(x_{0})P_{1}^{-1}P_{2}E_{2}(x_{0})E_{2}^{-1}(x_{0}+d_{2})\\
&\times P_{2}^{-1}P_{1}E_{1}(x_{0}+d_{2})=E_{1}^{-1}(x_{0})M(0)E_{1}(x_{0}) 
\end{eqnarray}
Then the resulting  transfer matrix becomes 
\begin{eqnarray}
\notag
M_{n} &=&M(0)M(d)M(2d)...M((n-1)d) \\*
\notag
&=&M(0)E_{1}^{-1}(d)M(0)E_{1}(d)E_{1}^{-1}(2d)M(0)E_{1}(2d)\\*
\notag
&&...E_{1}^{-1}((n-1)d)M(0)E_{1}((n-1)d)\\*
\notag 
&=&M(0)E_{1}^{-1}(d)M(0)E_{1}^{-1}(d)M(0)E_{1}^{-1}(d)\\*
\notag
&&... E_{1}^{-1}(d)M(0)E_{1}((n-1)d)\\*
%\notag 
&=&M(0)[E_{1}^{-1}(d)M(0)]^{n-1}E_{1}((n-1)d)  
\end{eqnarray}%

The transmission amplitudes $t_{\eta s_z}^{(n)}$ are found by solving the system of  equations
\begin{equation}
\Big(
1 \,\,,
r_{\uparrow }^{\uparrow } \,\,,
0 \,\,,
r_{\downarrow }^{\uparrow }%
\Big)^T%
=m%
\Big(
t_{\uparrow }^{\uparrow } \,\,, 
0 \,\,,
t_{\downarrow }^{\uparrow } \,\,, 
0\Big)^T%
\end{equation}%
where m is the total transfer matrix of  the barrier structure and $T$ denotes the transpose of the row vectors.
Then the transmission is given by $T=|t_{\eta s_z}^{(n)}|^2$.

All the results so far involve the four-component spinors (17). If we neglect the very small \cite{ezawa1,ezawa3} constant $\lambda_R$, the Hamiltonian (1) becomes block-diagonal 
and the eigenfunctions two-component spinors. This gives the simple analytic expressions (2), (4), (5).
\ \\

\end{document}